\documentclass{emulateapj}

\shortauthors{Draine \& Fraisse}
\shorttitle{Polarized FIR and Submillimeter Emission from Interstellar
  Dust}

\usepackage{epsfig}     

\newcommand{\Cabs}{C_\mathrm{abs}}
\newcommand{\Cabsa}{C_{\mathrm{abs},1}}
\newcommand{\Cabsb}{C_{\mathrm{abs},2}}
\newcommand{\Cpol}{C_\mathrm{pol}}
\newcommand{\Qpol}{Q_\mathrm{pol}}
\newcommand{\Qext}{Q_\mathrm{ext}}

\newcommand{\beq}{\begin{equation}}
\newcommand{\eeq}{\end{equation}}
\newcommand{\beqa}{\begin{eqnarray}}
\newcommand{\eeqa}{\end{eqnarray}}

\newcommand{\arctanh}{\mathrm{arctanh}}

\newcommand{\car}{\mathrm{car}}
\newcommand{\eff}{\mathrm{eff}}
\newcommand{\ext}{\mathrm{ext}}
\newcommand{\mrmax}{\mathrm{max}}
\newcommand{\obs}{\mathrm{(obs)}}
\newcommand{\pol}{\mathrm{pol}}
\newcommand{\sil}{\mathrm{sil}}

\newcommand{\dd}{\mathrm{d}}

\newcommand{\cm}{\mathrm{cm}}
\newcommand{\gm}{\mathrm{g}}
\newcommand{\um}{\mu\mathrm{m}}

\newcommand{\bB}{\mathbf{B}}
\newcommand{\bE}{\mathbf{E}}
\newcommand{\bk}{\mathbf{k}}
\newcommand{\bahat}{\mathbf{\hat{a}}}
\newcommand{\bxhat}{\mathbf{\hat{x}}}
\newcommand{\byhat}{\mathbf{\hat{y}}}
\newcommand{\bzhat}{\mathbf{\hat{z}}}

\begin{document}


\title{Polarized Far-Infrared and Submillimeter Emission from
  Interstellar Dust}
\author{Bruce T. Draine and Aur\'elien A. Fraisse}
\affil{Princeton University Observatory, Peyton Hall, Princeton, NJ
  08544, USA}
\email{draine@astro.princeton.edu}
\email{fraisse@astro.princeton.edu}
\submitted{Draft version September 7, 2008}


\begin{abstract}
Polarized far-infrared and submillimeter emission is calculated for
models of nonspherical dust grains that are constrained to reproduce
the observed wavelength-dependent extinction and polarization of
starlight.  For emission from regions where the magnetic field is
perpendicular to the line-of-sight, the far-infrared emission is
expected to have substantial linear polarization at wavelengths
$\lambda \gtrsim 100~\micron$, but the degree of linear polarization,
and its variation with wavelength, is model-dependent.  Models in
which the starlight polarization is produced by both amorphous
silicate and graphite grains have linear polarizations between $6\%$
and $10\%$ at $\lambda > 100~\micron$, but for some models in which
only silicate grains are spheroidal, the linear polarization increases
from about $3\%$ at $100~\um$ to about $15\%$ at 1~mm.  We briefly
discuss the implications of these results for removal of the polarized
dust emission from maps of the Cosmic Microwave Background, as well as
the possibility of discriminating among interstellar dust models based
on observations of far-infrared and submillimeter linear
polarization.
\end{abstract}




\section{Introduction}
\label{sec:intro}

The presence of dust grains in the interstellar medium has been
apparent since the recognition that obscuring material existed in at
least some regions of interstellar space~\citep{Barnard_1907,
  Barnard_1910, Trumpler_1930}.  Our knowledge of interstellar dust
has advanced considerably since then, but remains incomplete [see,
  e.g., the review by \citet{Draine_2003}, or the book by
  \citet{Whittet_2003}]. 

In the Milky Way, large fractions of abundant elements, such as Si and
Fe, appear to have been depleted from the gas phase
\citep{Jenkins_2004}, with the missing atoms believed to be present in
dust grains.  Studies of interstellar ``reddening'' produced by
wavelength-dependent extinction revealed broad spectroscopic features
at $9.8~\um$ and $18~\um$, attributed to amorphous silicate material
in submicron particles, and a strong extinction ``bump'' at
$2175$~\AA, usually attributed to aromatic carbon, perhaps graphite or
polycyclic aromatic hydrocarbons (PAHs), in particles with radii
$a \lesssim 0.01~\um$.  Another broad extinction feature, at
$3.4~\um$, is typical of C-H stretching modes in hydrocarbons.
Finally, the observed infrared emission from dust includes spectral
features at $3.3$, $6.2$, $7.7$, $8.6$, $11.3$ and $12.7~\um$ that can
be explained by a substantial population of particles with PAH
composition and sizes extending down to about $40$ atoms.

These spectroscopic features indicate that amorphous silicates and
carbonaceous materials, including PAHs, contribute a substantial
fraction of the interstellar grain mass, and models developed to try
to reproduce the observed extinction using observed abundances of the
elements invariably employ amorphous silicates and carbonaceous
materials as their main components. 

The serendipitous discovery \citep{Hiltner_1949, Hall_1949} of the
linear polarization of starlight demonstrated that interstellar grains
are both appreciably nonspherical and aligned with the interstellar
magnetic field, so that the dusty interstellar medium acts as a
polarizer.  Studies of starlight polarization in the ultraviolet
\citep{Clayton+Anderson+Magalhaes_etal_1992,
       Anderson+Weitenbeck+Code_etal_1996,
       Martin+Clayton+Wolff_1999} 
indicate that the smallest dust grains are apparently minimally
aligned \citep{Kim+Martin_1995}, and studies in the infrared [see,
  e.g., \citet{Whittet_2003}] find strong polarization in the
$9.8~\um$ Si-O stretch absorption feature, requiring that a
substantial fraction of the amorphous silicate particles be aligned.
Recent measurements of the polarization in the neighborhood of the
$3.4~\um$ C-H stretch absorption feature
\citep{Chiar+Adamson+Whittet_etal_2006} find only upper limits on
possible polarization associated with the feature, indicating that the
hydrocarbon material responsible for this feature is primarily in a
grain population that is not appreciably aligned, which may rule out
models where the hydrocarbon material is located in mantles coating
silicate cores.

While we know for sure that interstellar grains are not spherically
symmetric, their actual shapes are uncertain.  Because the degree of
alignment is unknown, we can only say that the grains must depart from
spherical symmetry sufficiently to produce the observed polarization.

Interstellar grains radiate strongly in the far-infrared.  Because the
grains are nonspherical and partially alig\-ned, this emission must be
partially polarized.  The polarization of the far-infrared emission
has been measured for infrared-bright regions in star-forming
molecular clouds \citep[see the review by][]{Hildebrand_2005}.  The
degree of polarization of the thermal emission from dust in these
regions is observed to have a complicated wavelength dependence, with
the polarization at $350~\um$ observed to be significantly smaller
than the $100~\um$ and $850~\um$ polarization
\citep{Vaillancourt_2002, 
       Hildebrand+Kirby_2004, 
       Hildebrand_2005,
       Vaillancourt_2007},
which has been interpreted as indicating two separate populations of
emitting grains in these clouds.

The polarization of the large-scale submillimeter emission from dust
in the interstellar medium of the Milky Way has now been measured by
several experiments, including the Archeops balloon
\citep{Benoit+Ade+Amblard_etal_2004,
       Ponthieu+Macias-Perez+Tristram_etal_2005}
and the {\it WMAP} satellite
\citep{Page+Hinshaw+Komatsu_etal_2007,
       Gold+Bennett+Hill_etal_2008}.
A quantitative understanding of this emission is important for removal
of polarized Galactic ``foregrounds" from maps of the cosmic microwave
background (CMB), but also offers an opportunity to test models for
interstellar grains.

In this paper, we construct models of interstellar dust that are
consistent with the observed wavelength-dependent interstellar
extinction, and the observed linear dichroism of the interstellar
medium.  We will calculate the polarized infrared and submillimeter
emission from different models of interstellar dust heated by the
general interstellar starlight, and, for each of them, compute the
corresponding expected percentage polarization.  This will allow us to
gain a sense of the degree to which the predicted far-infrared
polarization is model-dependent.

We discuss the observed wavelength-dependent extinction and
polarization of starlight in \S\ref{sec:obs}, and the basics of the
dust models we consider in \S\ref{sec:models}.
\S\ref{sec:cross_sections} provides a description of  the technique we
use to compute optical cross sections, while \S\ref{sec:picket_fence}
introduces the formalism associated with ``picket-fence" alignment.
In \S\ref{sec:size_dist_align_func}, we present the size distributions
and alignment functions that provide the best fit to the observed
extinction and polarization discussed in \S\ref{sec:obs}.  We describe
the formalism we use to calculate the partially-polarized far-infrared
emission by interstellar dust in \S\ref{sec:formalism}.  The
corresponding results are presented and discussed in
\S\ref{sec:ir_em}.  Finally, \S\ref{sec:summary} offers a summary of
the main results of this paper, and a brief discussion of their
consequences for studies of the CMB.


\section{Extinction and Polarization of Starlight}
\label{sec:obs}

The grain models considered here will be required to reproduce the
observed extinction by interstellar dust.  Gas and dust are generally
well-mixed, and the mixture is characterized by the observed
extinction cross section at wavelength $\lambda$ per H nucleon
\beq
\sigma_\ext^\obs(\lambda) \equiv
0.4\,(\ln 10)\, \frac{A_\lambda/{\rm mag}}{N_{\rm H}},
\eeq
where $A_\lambda$ is the measured extinction, and $N_{\rm H}$ the
column density of H nucleons.  In diffuse sightlines, $N_{\rm H}$ can
be determined from ultraviolet observations of Lyman-$\alpha$
absorption by atomic H and Lyman- and Werner-band absorption from
H$_2$. The extinction produced by dust in the diffuse interstellar
medium has been studied from about $0.1~\micron$ to slightly over
$10~\micron$ [see, e.g., the review by \citet{Fitzpatrick_2004}].  At
wavelengths  $4 \lesssim \lambda \lesssim 8~\micron$, the extinction
is somewhat controversial
\citep{Lutz+Feuchtgruber+Genzel_etal_1996,
       Rosenthal+Bertoldi+Drapatz_2000,
       Indebetouw+Mathis+Babler_etal_2005,
       Flaherty+Pipher+Megeath_etal_2007},
but the present study will be limited to the wavelength range
$0.1 < \lambda < 2.65~\micron$, where the extinction by dust in
diffuse regions appears to be securely known.  We take the wavelength
dependence of the observed extinction to be given by the
\citet{Fitzpatrick_1999} parametrization for
$R_V \equiv A_V/(A_B-A_V) = 3.1$, and, following
\citet{Bohlin+Savage+Drake_1978}, the value of $N_{\rm H}/(A_B-A_V)$
is set to $5.8\times 10^{21}~\cm^{-2}/{\rm mag}$.

The light emitted by stars is usually unpolarized, and the observed
fractional polarization is produced by linear dichroism of the
interstellar medium -- a difference in the attenuation between two
orthogonal linear polarization modes.  We define the fractional
polarization
\beq
p \equiv \frac{F_1-F_2}{F_1+F_2} = 
\frac{e^{-\tau_1} - e^{-\tau_2}}{e^{-\tau_1} + e^{-\tau_2}} = 
\tanh(\Delta\tau),
\eeq
where $\Delta\tau \equiv (\tau_2-\tau_1)/2$, $F_1$ and $F_2$ are the
fluxes in each of two orthogonal polarization modes, and $\tau_1$ and
$\tau_2$ the extinctions for these modes.  The observed polarization
fraction $p(\lambda)$ tends to peak at a wavelength $\lambda_\mrmax$
between about $0.5$ and $0.9~\micron$, with the larger values of
$\lambda_\mrmax$ coming from sightlines through dense clouds; diffuse
cloud sightlines tend to have $\lambda_\mrmax \approx 0.55~\micron$
\citep{Whittet_2003}.

The linear dichroism is characterized by a polarization cross section
at wavelength $\lambda$ per H nucleon
\beq
\sigma_\pol^\obs(\lambda) \equiv 
\frac{\Delta\tau}{N_{\rm H}} =
\frac{1}{N_{\rm H}}\,\arctanh[p(\lambda)],
\eeq
which, for $p \ll 1$, can be approximated as
\beq
\sigma_\pol^\obs(\lambda) \approx \frac{p(\lambda)}{N_{\rm H}}.
\eeq
The strength of the polarization is characterized by the polarization
per unit extinction $p_\mrmax/A(\lambda_\mrmax)$, where 
$p_\mrmax = p(\lambda_\mrmax)$.  On some sightlines the polarization
per unit extinction is as large as  
$p_\mrmax/A(\lambda_\mrmax) \approx 0.03/{\rm mag}$
\citep{Whittet_2003}.  Smaller values of  $p_\mrmax/A(\lambda_\mrmax)$
are presumed to arise where the local magnetic field is not orthogonal
to the sightline, where the magnetic field direction varies along the
line-of-sight, or where the processes responsible for grain alignment
are for some reason less effective.  A successful grain model must be
able to account for the largest observed values, i.e.,
$p_\mrmax/A(\lambda_\mrmax) \approx 0.03/{\rm mag}$.

Various approaches have been taken to describing the wavelength
dependence of interstellar polarization
[see, e.g.,\citet{Martin+Clayton+Wolff_1999}]. 
Following \citet{Draine+Allaf-Akbari_2006}, we take it to be given by
the empirical ``Serkowski law'' \citep{Serkowski_1973}
\beq
p(\lambda) \approx p_\mrmax\,
\exp\left[-K\left(\ln\frac{\lambda}{\lambda_\mrmax}\right)^2\right]
\eeq
for $0.15 < \lambda < 1.39~\micron$, with
$\lambda_\mrmax = 0.55~\micron$ and $K = 0.92$.  For
$1.39 < \lambda < 5~\micron$, we assume the power-law dependence
$p(\lambda) \propto \lambda^{-1.7}$, consistent with the observations
by \citet{Martin+Adamson+Whittet_etal_1992}.


\section{Dust Models}
\label{sec:models}

As in previous studies 
\citep{Weingartner+Draine_2001,
       Draine+Allaf-Akbari_2006},
we model the dust as a mixture of carbonaceous particles (including
PAHs) and amorphous silicate grains.  Models of the infrared emission
from interstellar dust \citep{Li+Draine_2001, Draine+Li_2007} require
a population of very small PAH particles, containing 
$\mathrm{C}/\mathrm{H} \approx 55$~ppm, or about $22\%$ of all of the
C atoms if the C abundance in the ISM is equal to the the current best
estimate for the solar abundance
$(\mathrm{C}/\mathrm{H})_\odot = (247 \pm 28)$~ppm
\citep{Asplund+Grevesse+Sauval_etal_2005}.  We assume the PAHs to have
the absorption cross sections given by \citet{Draine+Li_2007}, with a
strong absorption feature at $2175$~\AA, and to contribute
$\sigma_\ext(\mathrm{PAH},\lambda)$ to the overall extinction.  The
PAH population is expected to be randomly-oriented
\citep{Lazarian+Draine_1999}, and therefore is assumed to contribute
no polarization.  This is consistent with the observed weakness or
absence of polarization in the aforementioned $2175$~\AA~feature
\citep{Clayton+Anderson+Magalhaes_etal_1992,
       Wolff+Clayton+Kim_etal_1997}.

In addition to the PAHs, the model includes populations of larger
silicate particles and carbonaceous grains.  The size distribution of
these particles is a priori unknown, as is the degree of alignment.
To determine these quantities, we will follow the approach pioneered
by \citet{Kim+Martin_1995}, seeking to find a model grain size
distribution and size-dependent degree of alignment that can
simultaneously reproduce both the wavelength-dependent extinction
cross section $\sigma_\ext(\lambda)$ and the wavelength-dependent
polarization $\sigma_\pol(\lambda)$.

In our models, the dust grains will be assumed to be oblate spheroids,
with semimajor axis $a$ along the symmetry axis $\bahat$, and $b > a$
perpendicular to this axis.  For a given axial ratio $b/a$, we
characterize the grain size by the effective radius
$a_\eff \equiv (ab^2)^{1/3}$, which corresponds to the radius of an
equal-volume sphere.


\section{Optical Cross Sections}
\label{sec:cross_sections}

\begin{figure*}
\begin{center}
\epsfig{file=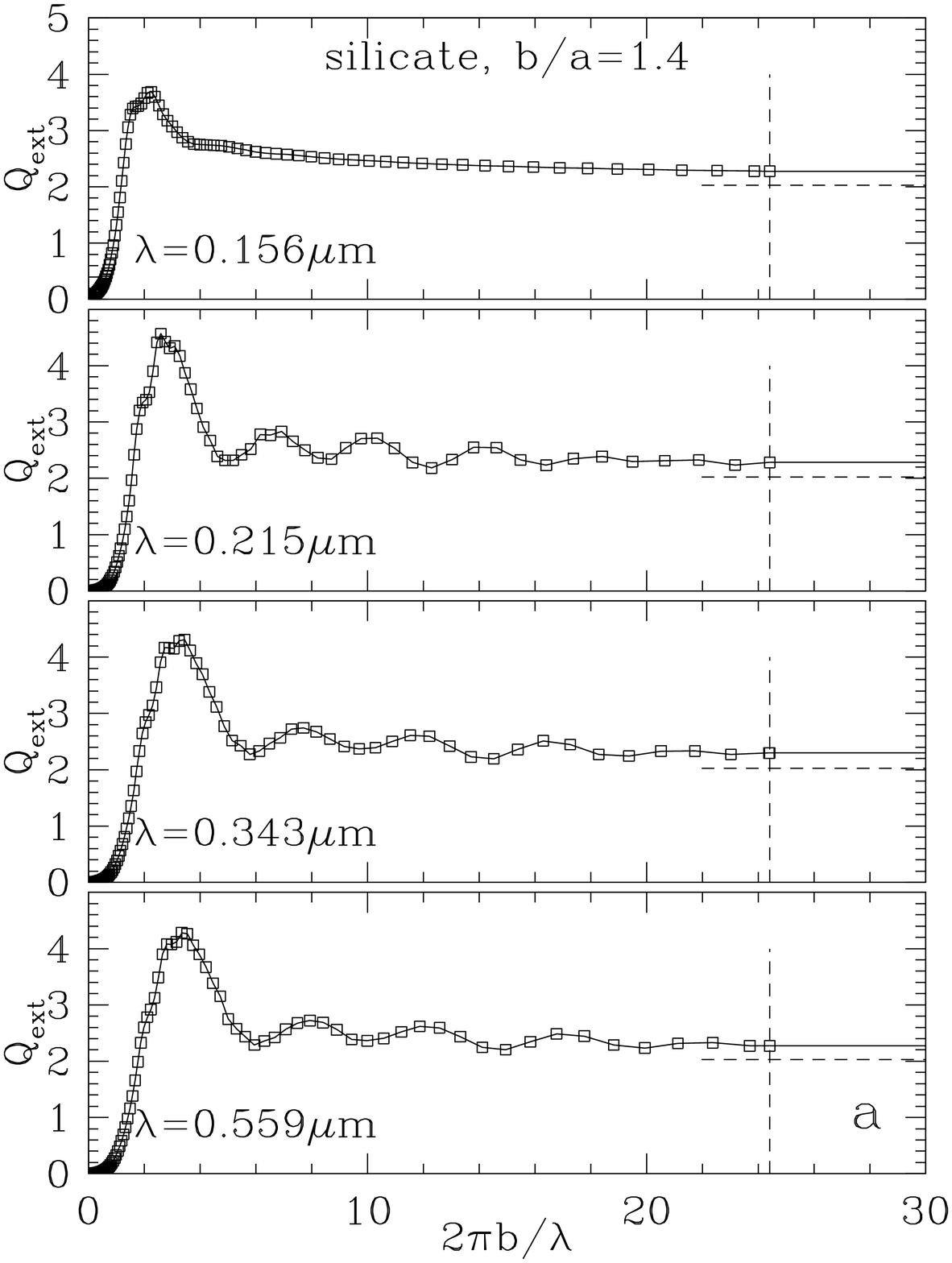,
        width=6.0cm,
	angle=0}
\epsfig{file=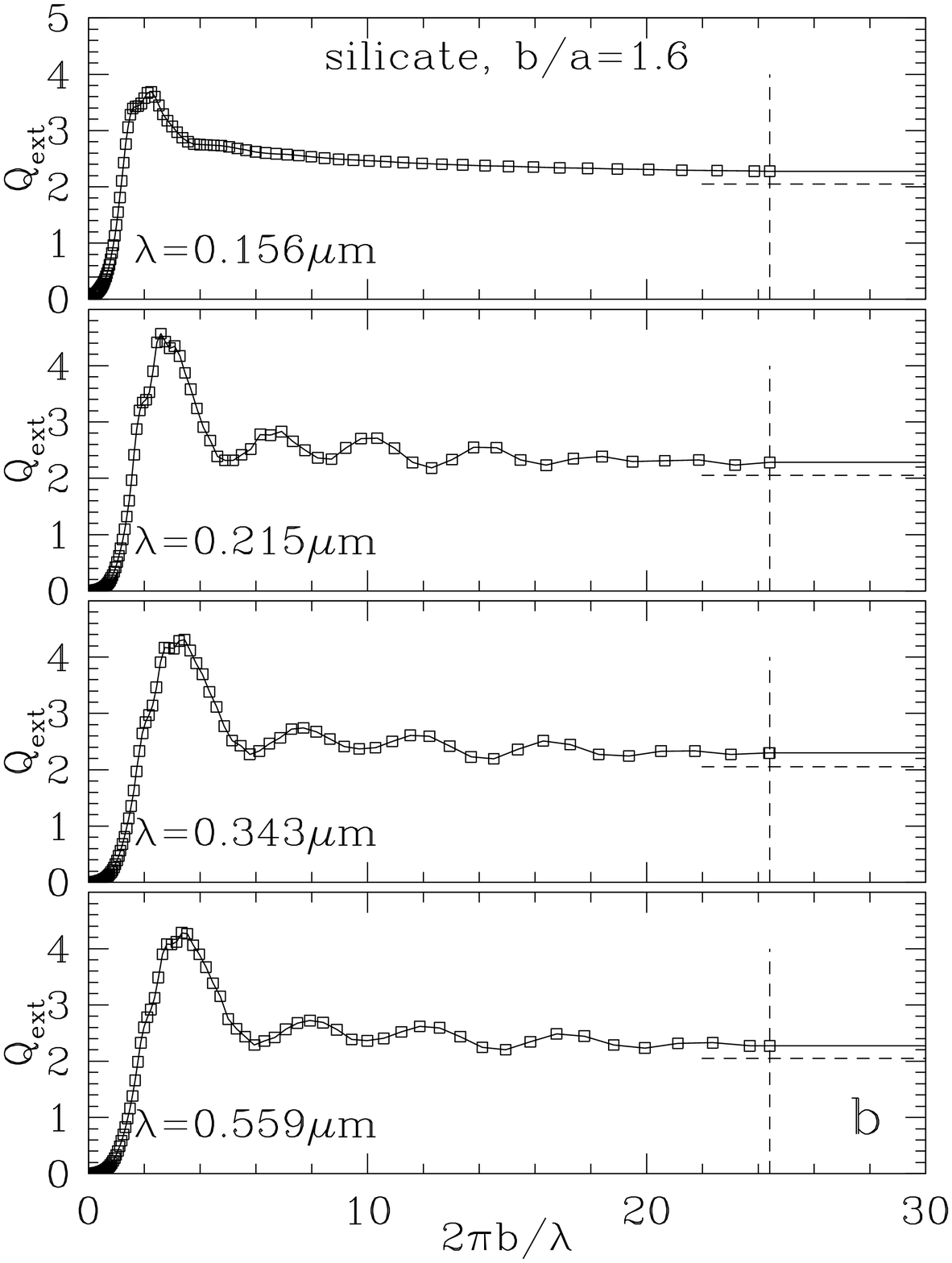,
        width=6.0cm,
        angle=0}
\epsfig{file=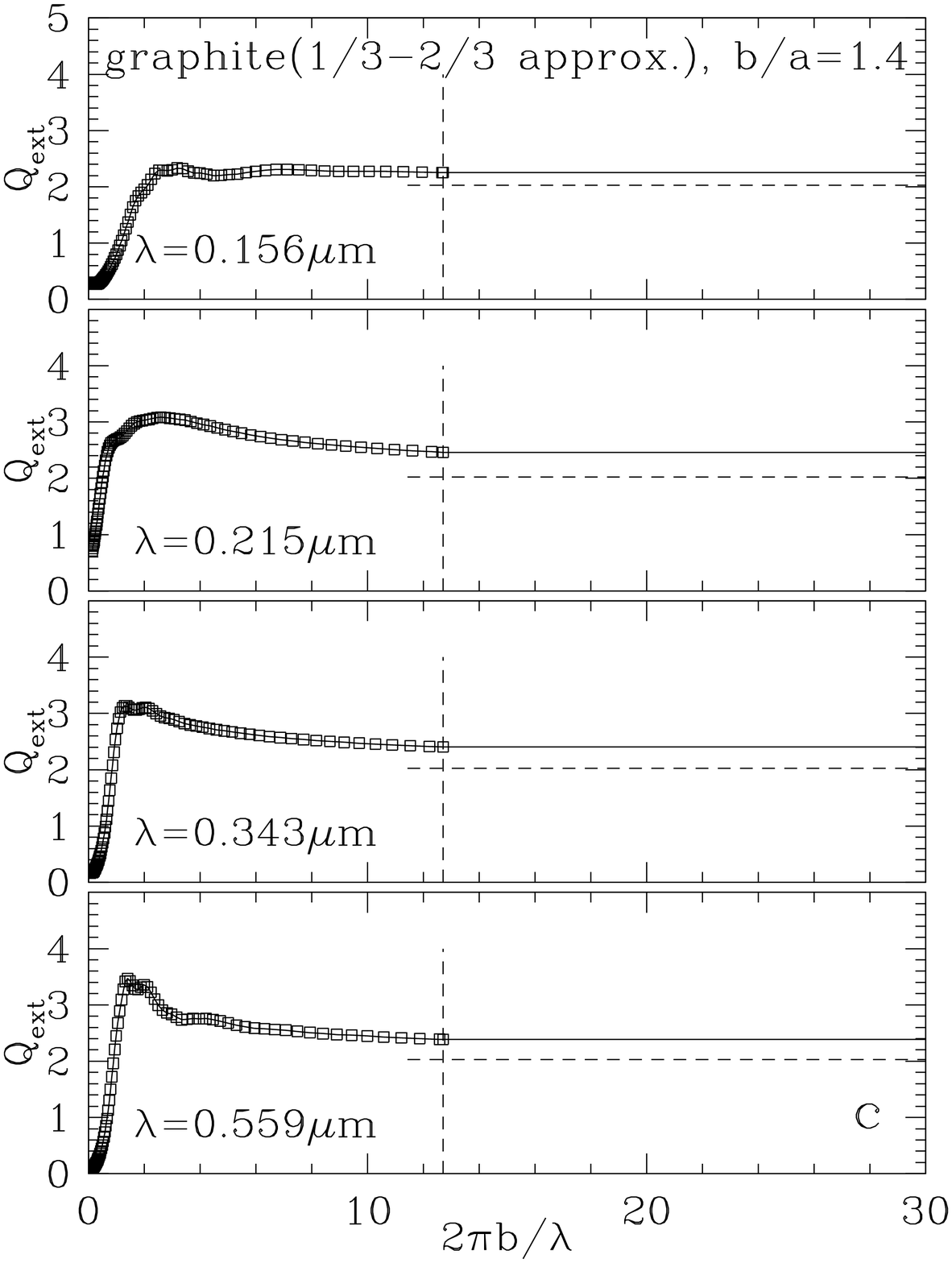,
        width=6.0cm,
	angle=0}
\epsfig{file=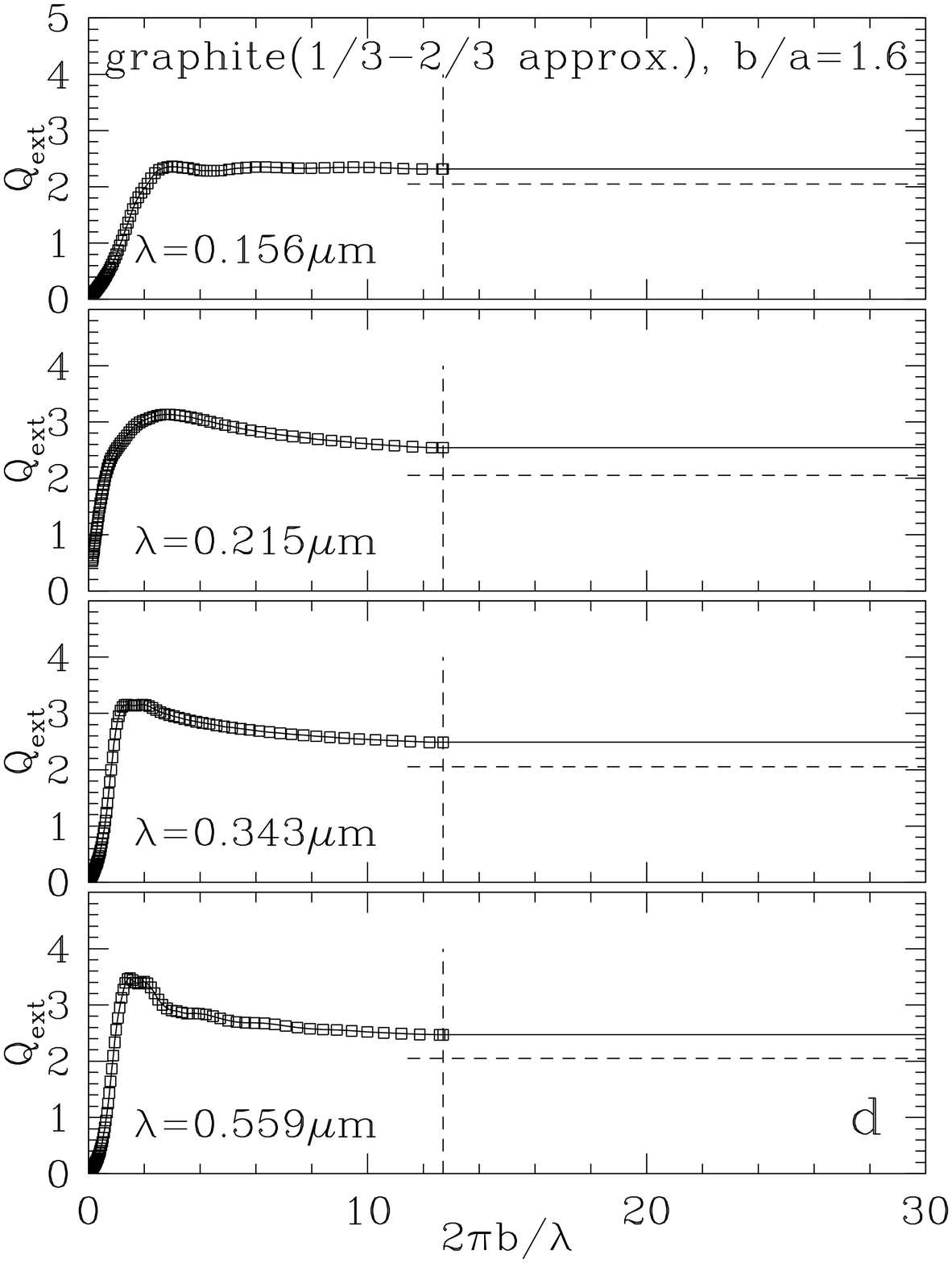,
        width=6.0cm,
        angle=0}
\caption{\label{fig:qext_v_b/lambda} $\Qext(b/\lambda,b/a,\lambda)$
	 vs. $2\pi b/\lambda$ at four wavelengths for
	 randomly-oriented silicate ({\it top}) and graphite
	 ({\it bottom}) spheroids with axial ratios $b/a$ of 1.4 and
	 1.6.  The vertical dashed lines show the maximum value of 
	 $2\pi b/\lambda$ allowed by the EBCM calculations with the
	 si\-li\-cate di\-elec\-tric function.  Horizontal dashed
	 lines indicate 
	 $Q_{\rm ext}(b/\lambda \to \infty,b/a,\lambda)$ in each
	 case.}
\end{center}
\end{figure*}

\begin{figure*}
\begin{center}
\epsfig{file=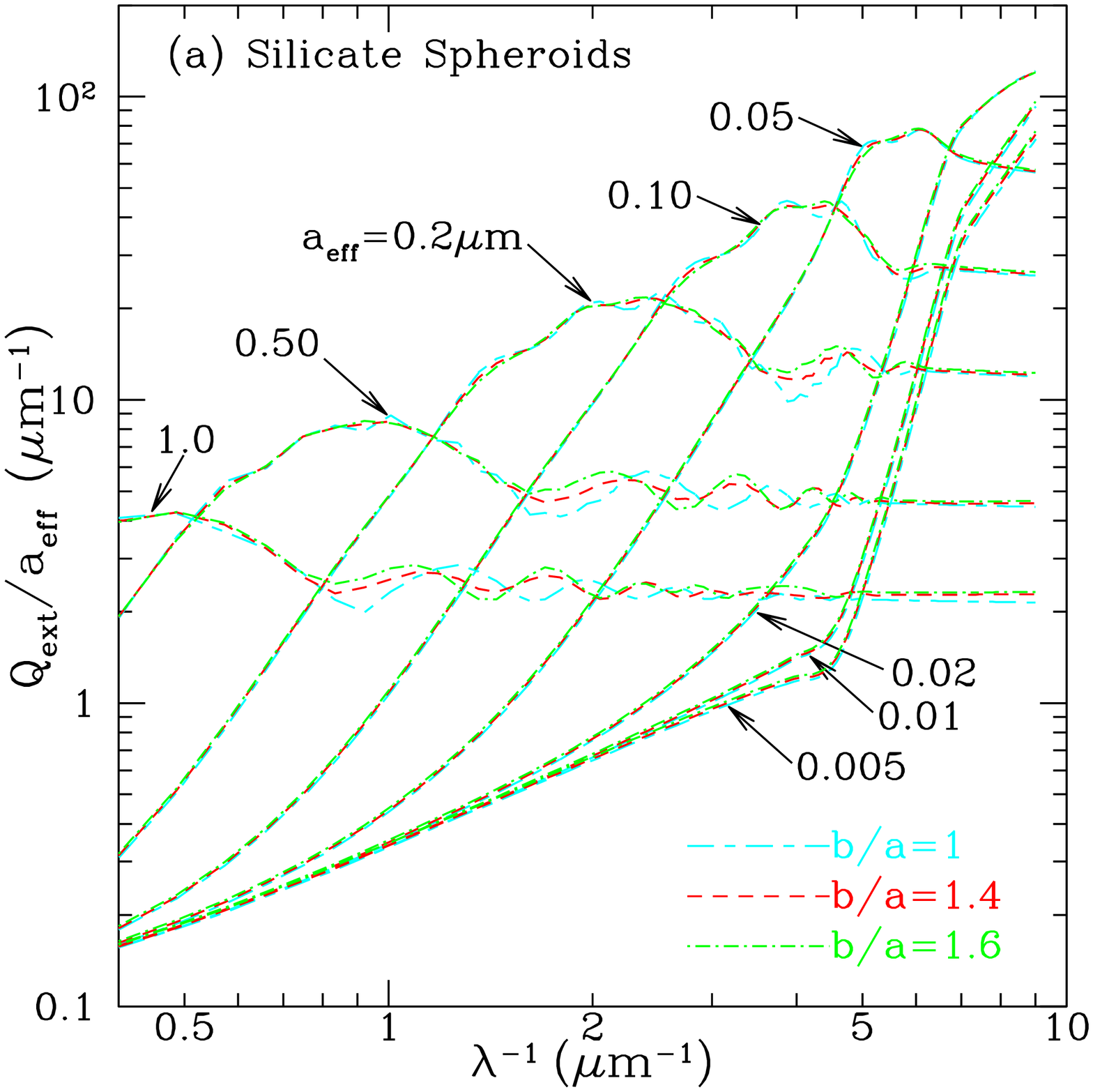,
        width=7.2cm,
	angle=0}
\epsfig{file=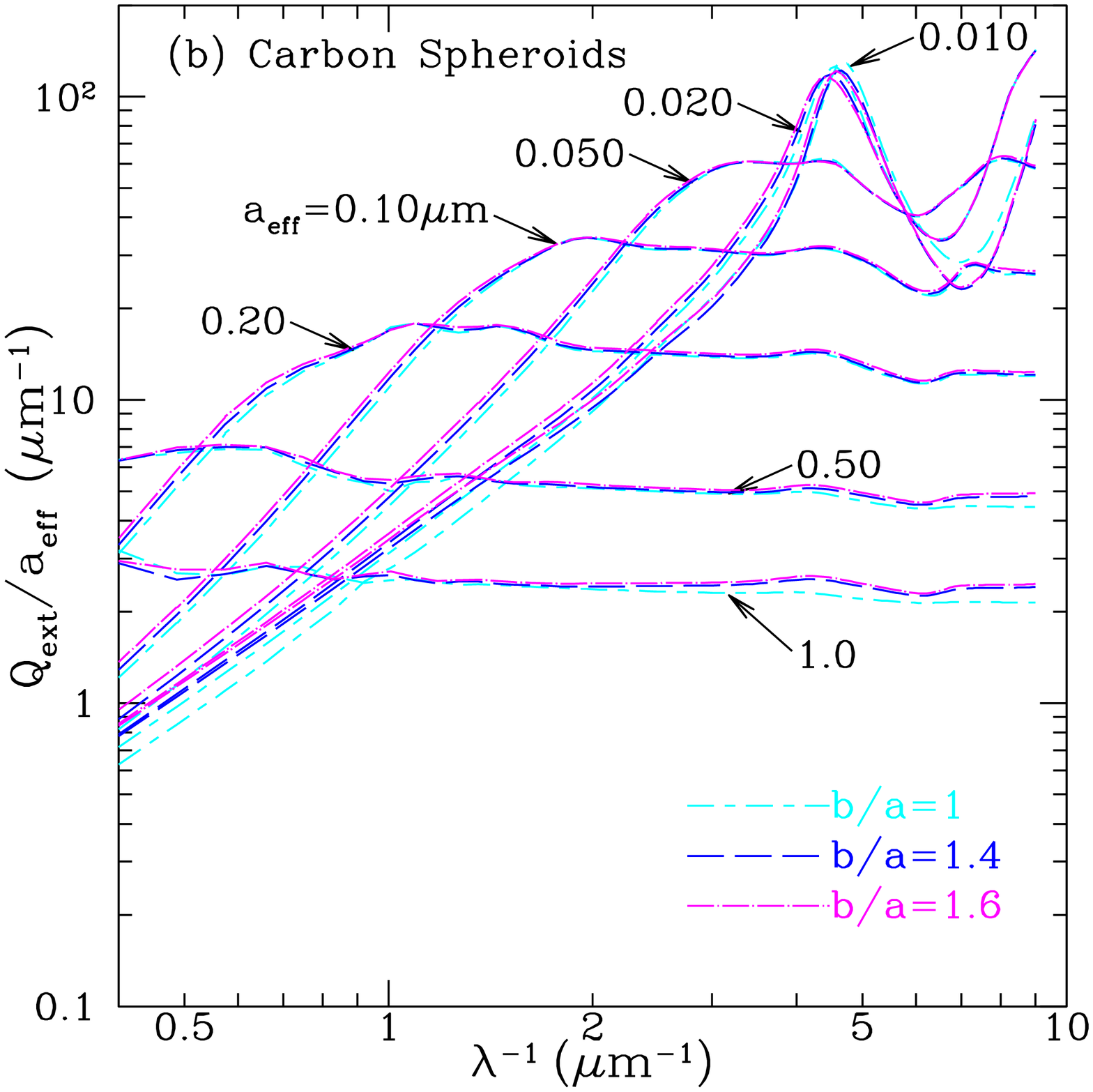,
        width=7.2cm,
        angle=0}
\caption{\label{fig:qext vs a} Extinction cross sections
	 $\Qext(a_\eff,b/a,\lambda)/a_\eff$ vs. inverse wavelength
	 $1/\lambda$ for ({\it left}) silicate, and ({\it right})
	 graphite spheres and spheroids with axial ratios of
	 $b/a = 1.4$ and $b/a = 1.6$.  Curves are labeled by
	 $a_\eff~(\mathrm{in~}\micron)$.}
\end{center}
\end{figure*}

For a nonspherical grain of solid volume $V$, we define the extinction
efficiency factor 
\beq
Q_\ext \equiv \frac{C_\ext}{\pi a_\eff^2},
\eeq
where $C_\ext$ is the extinction cross section, and $a_\eff$ the
radius of an equal-volume sphere, i.e., $(3V/4\pi)^{1/3}$.  The
extinction cross sections for polarized light are calculated for
oblate spheroids using the ``extended boundary condition method''
(EBCM) introduced by \citet{Waterman_1971} and developed by
\citet{Mishchenko+Travis_1994} and
\cite{Wielaard+Mishchenko+Macke_etal_1997}.  We use the open-source
code {\tt ampld.lp.f} made available by \citet{Mishchenko_2000}.  EBCM
codes encounter numerical difficulties when the target becomes large
compared to the wavelength.  For the silicate oblate spheroids
($b/a > 1$), we find that {\tt ampld.lp.f} sometimes fails for
$b/\lambda > 3.885$.  Thus, for $b/\lambda > 3.885$, and for both
$b/a = 1.4$ and $b/a = 1.6$ (values we will focus on from now on), we
take 
\beq
\label{eq:Qext_for_silicate_for_large_b/lambda}
Q_{\rm ext}[b/\lambda,b/a,\lambda] \approx 
Q_{\rm ext}[3.885,b/a,\lambda]
\eeq
for silicates.  The EBCM assumes an isotropic dielectric function, but
graphite is a highly anisotropic material.  The eigenvalues of the
graphite dielectric tensor are
$(\epsilon_\parallel,\epsilon_\perp,\epsilon_\perp)$.  In order to
estimate $Q_{\rm ext}$ and $Q_\pol$ for spheroids composed of
graphite, we use the so-called ``1/3-2/3 approximation'': we assume
that 1/3 of the spheroids have an isotropic dielectric function
$\epsilon = \epsilon_\parallel$, while 2/3 of them have
$\epsilon = \epsilon_\perp$.  This gives the correct cross sections in
the ``electric dipole limit'' $a_\eff \ll \lambda$, and
\citet{Draine+Malhotra_1993} used the discrete dipole approximation to
show that the 1/3-2/3 approximation is reasonably accurate even for
finite grain size.  The actual composition of the carbonaceous grains
is unknown.  Based on the profile of the $3.4~\um$ C-H stretching
mode, \citet{Pendleton+Allamandola_2002} estimate that about $85\%$ of
the carbon is in aromatic structures.  We do not expect interstellar
grains to be single-crystal graphite, but it is hoped that the
approach adopted here will provide cross sections that will
approximate the carbonaceous material present in the interstellar
medium.

The dielectric function for graphite is somewhat more extreme in the
ultraviolet than that for silicate.  For the graphite oblate
spheroids, we find that {\tt ampld.lp.f} sometimes fails for
$b/\lambda > 2.023$.  Thus, for graphite grains with 
$b/\lambda > 2.023$, we take
\beq
\label{eq:Qext_for_graphite_for_large_b/lambda}
Q_{\rm ext}[b/\lambda,b/a,\lambda] \approx
Q_{\rm ext}[2.023,b/a,\lambda].
\eeq

The top panel of Figure \ref{fig:qext_v_b/lambda} shows $Q_\ext$ as a
function of $2\pi b/\lambda$ at four se\-lec\-ted wave\-lengths for
randomly-oriented\footnote{Random orientation is approximated by
  averaging over the three independent orientations of spheroids in
  the ``picket-fence'' approximation described in
  \S\ref{sec:picket_fence}.}
silicate spheroids with $b/a = 1.4$ and $b/a = 1.6$.  In the limit
$b/a \to \infty$, we expect $Q_{\rm ext}\,\pi a_\eff^2$ to approach
twice the geometric cross section, with 50\% of the total extinction
produced by rays that impinge on the grain, and 50\% contributed by
small angle scattering resulting from diffraction [see, e.g., the
  discussion of the ``extinction paradox'' in
  \citet{Bohren+Huffman_1983}].
If (1/3,1/3,1/3) of the spheroids have symmetry axis
$(\bahat\parallel\bxhat,\bahat\parallel\byhat,\bahat\parallel\bzhat)$,
the extinction efficiency factor for light propagating in the
$\bxhat$, $\byhat$, or $\bzhat$ directions is such that
\beq
Q_{\rm ext}(b/\lambda \to \infty) =
\frac{2}{3}\left[(b/a)^{2/3} + 2(a/b)^{1/3}\right].
\eeq  
The horizontal dashed line shows this limiting value; we see that the
calculated cross sections appear to be converging to this limit.  Our
assumption summarized by
equation~(\ref{eq:Qext_for_silicate_for_large_b/lambda}) will
overestimate $Q_{\rm ext}$ slightly.  However, for realistic size
distributions, the extinction at $\lambda > 0.1~\micron$ is generally
dominated by grains with $b/\lambda \lesssim 1$. 

The bottom panel of Figure~\ref{fig:qext_v_b/lambda} is the same as
the top panel, but for graphite spheroids.  Note that while
$Q_{\rm ext}$ for the silicate spheroids reaches values as large as
4.7 (for $\lambda = 0.215$~\AA~in the top panel), the carbon grains
reach only $Q_{\rm ext} \approx 3.1$ for $b/a = 1.4$, and
$Q_{\rm ext} \approx 2.8$ for $b/a = 1.6$; the refractive index
matters!  We also see that the extinction approaches twice the
geometric cross section for $b/\lambda \gtrsim 2$, so we may use
equation~(\ref{eq:Qext_for_graphite_for_large_b/lambda}) to estimate 
$Q_{\rm ext}$ for large values of $b/\lambda$.

Finally, Figure \ref{fig:qext vs a} shows extinction cross sections as
a function of wavelength for silicate and carbon spheroids of selected
values of the effective radius $a_\eff$.


\section{Picket-Fence Alignment}
\label{sec:picket_fence}

\begin{figure*}
\begin{center}
\epsfig{file=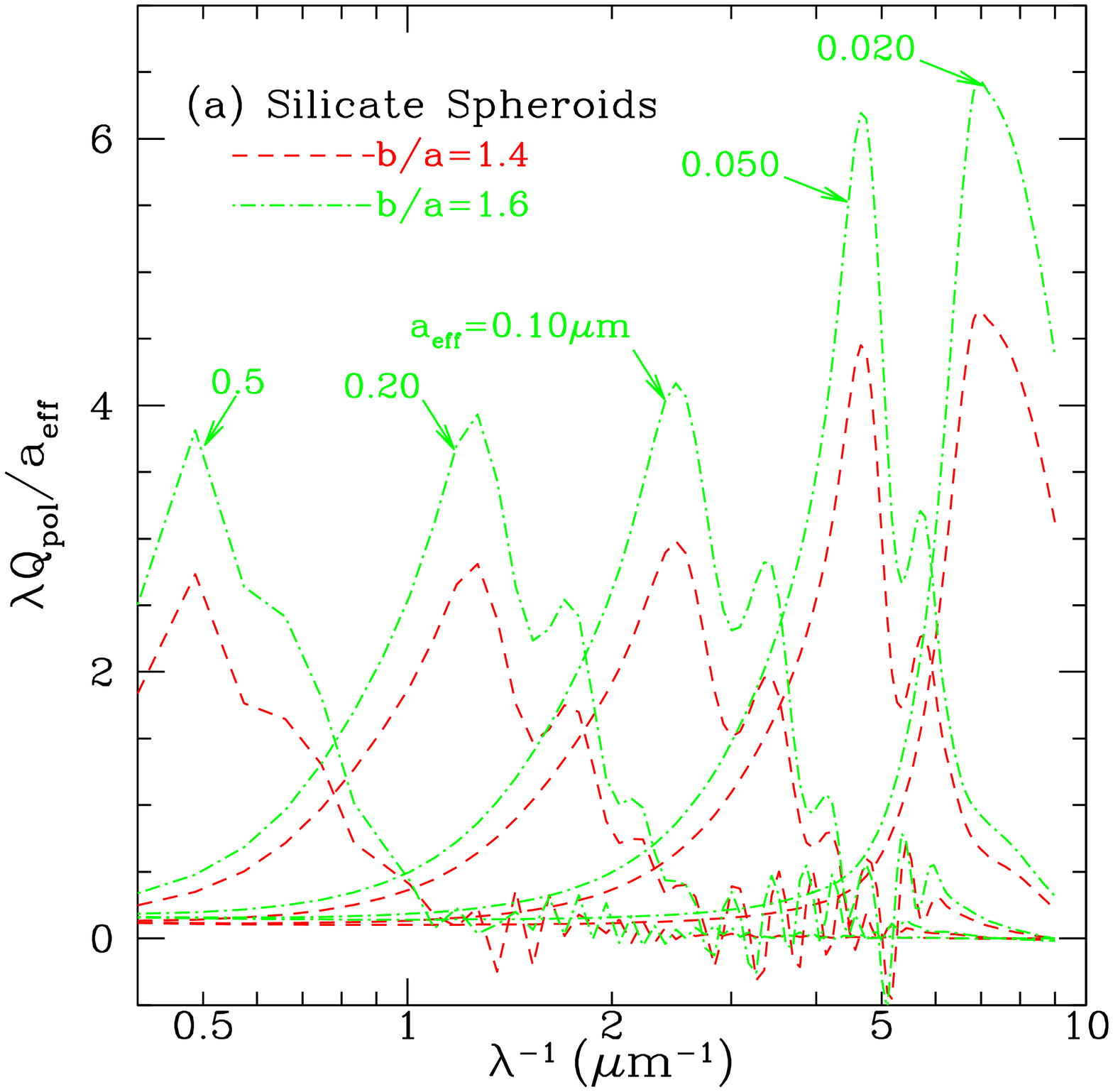,
        width=7.2cm,
	angle=0}
\epsfig{file=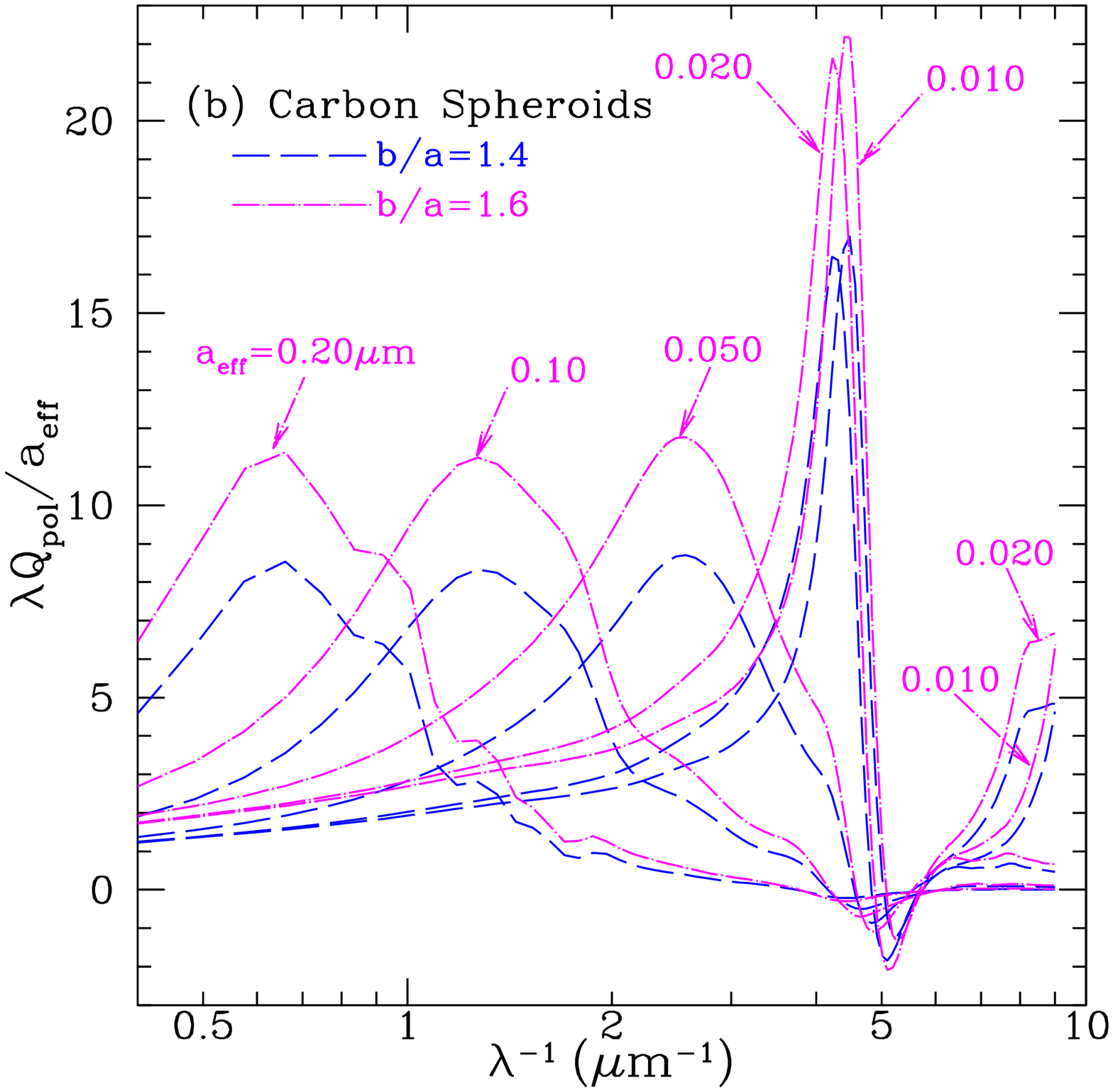,
        width=7.2cm,
        angle=0}
\caption{\label{fig:qpol vs a} $\lambda\,\Qpol/a_\eff$ vs. inverse
	 wavelength for ({\it left}) silicate spheroids, and
	 ({\it right}) graphite spheroids with axial ratios 
	 $b/a = 1.4$ and $b/a = 1.6$.  Curves are labeled by 
	 $a_\eff~(in~\micron)$.  The pronounced drop in $Q_\pol$ for
	 graphite grains as $\lambda$ increases from $0.20~\micron$ to
	 $0.22~\micron$ is a result of the wavelength dependence of
	 the dielectric function $\epsilon_\perp$ (see text).}
\end{center}
\end{figure*}

The process of interstellar grain alignment 
[see, e.g., \citet{Roberge_2004}] leads to the short axis of grains
(the axis with the largest moment of inertia) tending to align with
the magnetic field direction.  For oblate spheroids, the symmetry axis
$\bahat$ therefore aligns with the magnetic field, and maximum
polarization of starlight occurs for sightlines perpendicular to the
magnetic field direction.

We adopt a coordinate system with the line-of-sight in the $\bzhat$
direction, and the magnetic field in the $\bxhat$ direction.  For
simplicity, we will assume ``picket-fence'' alignment, i.e., that a
fraction $(1+2f)/3$ of the oblate spheroids are oriented with the
symmetry axis $\bahat$ in the $\bxhat$ direction, a fraction $(1-f)/3$
with $\bahat \parallel \byhat$, and a fraction $(1-f)/3$ with
$\bahat \parallel \bzhat$.  Thus, $f = 0$ corresponds to random
orientation (1/3 of the grains in each orientation), and $f = 1$ to
perfect alignment (all of the grains with $\bahat \parallel \bxhat$).

``Picket-fence'' alignment is not physically realistic.  Interstellar
grains will have a continuous distribution of orientations, and a
grain with angular momentum that is not aligned with $\bB$ will
precess around~$\bB$.  The advantage of assuming picket-fence
alignment is that for each grain size and wavelength, the extinction
cross section needs to be calculated for only 3 grain orientations.
We expect that picket-fence alignment with alignment fraction $f$ will
closely reproduce the polarization produced by grains with a
continuous distribution of alignments provided 
$f=(3/2)\left[\langle (\bahat\cdot \bxhat)^2 \rangle - (1/3)\right]$.

Let $C_{\ext,j}(m,a_\eff,b/a,\lambda)$ be the extinction cross section
at wavelength $\lambda$ for a spheroid of material $m$, with effective
radius $a_\eff$ and axial ratio $b/a$, where $j$ refers to the
orientation of the grain relative to incident polarized light.  We
discretize the size range into $N$ size bins $a_{{\rm eff},i}$ (with
$i\in [1,N]$), uniformly spaced in $\log(a_{{\rm eff},i})$, with
$n_{\sil,i}$ and $n_{\car,i}$ the numbers of silicate and carbonaceous
grains per H nucleon, and $f_i$ the fractional alignment of grains of
size $a_{{\rm eff},i}$.  The process of interstellar grain alignment
remains uncertain, and it is possible that the alignment function $f$
might depend on grain composition as well as on grain size.  In this
paper, we will assume that $f_i$ depends only on grain size, not on
the material.

The ``average'' extinction curve we will fit to is an average over
sightlines making various angles to the magnetic field direction, so
it is appropriate to compare the observed extinction with the
extinction calculated for randomly-oriented spheroids.
\begin{widetext}
\noindent We define the dimensionless efficiency factor for extinction
by randomly-oriented grains through
\beq
Q_\ext\!\left(m,\frac{a_\eff}{\lambda},\frac{b}{a}\right) \equiv 
\frac{1}{\pi a_\eff^2} \frac{1}{3} \sum_{j=1}^{3}
C_{\ext,j}\!\left(m,a_\eff,\frac{b}{a},\lambda\right)\!.
\eeq
The index $j$ runs from $1$ to $3$ as we require three orientations
for picket-fence alignment with oblate spheroids: 
\beqa
j &=&1\mathrm{~for~}\bk \perp \bahat
      \mathrm{~and~}
      \bE\hspace{0.049cm}\parallel\hspace{0.049cm}\bahat,\nonumber\\
j &=&2\mathrm{~for~}\bk \perp \bahat
      \mathrm{~and~}
      \bE \perp \bahat,\\
j &=&3\mathrm{~for~}\bk\hspace{0.049cm}\parallel\hspace{0.049cm}\bahat
      \mathrm{~and~} \bE \perp \bahat\nonumber,
\eeqa
where $\bk$ is the incident wavevector.  We also define an efficiency
factor for polarization by perfectly-aligned grains: 
\beq
Q_\pol\!\left(m,\frac{a_\eff}{\lambda},\frac{b}{a}\right) \equiv 
\frac{1}{\pi a_\eff^2} \frac{1}{2}
\left[C_{\ext,2}\!\left(m,a_\eff,\frac{b}{a},\lambda\right) -
      C_{\ext,1}\!\left(m,a_\eff,\frac{b}{a},\lambda\right)\right]\!.
\eeq
Figure \ref{fig:qpol vs a} shows the polarization efficiency factors
$Q_{\rm pol}$ as a function of wavelength $\lambda$ for selected sizes
of silicate and carbon spheroids.  Note that $Q_{\rm pol}$ tends to
peak near $2\pi a_\eff/\lambda \approx 1.6$ for the silicate
spheroids, whereas it peaks near~0.8 for the carbon spheroids.  Small
carbon spheroids show a strong peak in $Q_{\rm pol}$ near
$\lambda \approx 0.23~\um$; this is a result of the strong peak in
Im($\epsilon_\perp$) due to $\pi \rightarrow \pi^*$ electronic
transitions near this frequency.

The model we consider here involves a mixture of silicate and carbon
(including PAHs) dust grains.  In such a case, the extinction cross
section per H nucleon is given by
\beq
\sigma_\ext(\lambda) = 
\sigma_\ext({\rm PAH},\lambda) +
\sum_m \sum_{i=1}^N
n_{m,i}\,\pi a_{{\rm eff},i}^2\,
Q_\ext\!\left[m,\frac{a_{{\rm eff},i}}{\lambda},
              \left(\frac{b}{a}\right)_{{\!}m}\,\right]\!,
\eeq
where we sum over materials $m = \{\sil,\car\}$ and over grain size
$a_{{\rm eff},i}$, with $n_{m,i}$ the number of grains of composition
$m$ and size $a_{{\rm eff},i}$ per H nucleon.  The polarization cross
section for the model is
\beq
\sigma_\pol(\lambda) =
\sum_m \sum_{i=1}^N f_i\, n_{m,i}\, \pi a_{{\rm eff},i}^2\,
Q_\pol\!\left[m,\frac{a_{\eff,i}}{\lambda},
              \left(\frac{b}{a}\right)_{{\!}m}\,\right]\!.
\eeq

A priori, we do not know the grain shape, except that the grains must
be sufficiently nonspherical to be able to reproduce the observed
polarization of starlight.  We will proceed using trial grain shapes,
and will see whether we are able to reproduce the polarization with a
size-dependent alignment fraction $f(a) \leq 1$.

For a given model (with assumed axial ratio $(b/a)_m$ for material
$m$), we use the technique described in \S\ref{sec:cross_sections} to
obtain the cross sections $C_{\ext,j}[m,a_{\eff,i},(b/a)_m,\lambda]$,
and then seek the abundances $n_{i,m}$ and alignment function $f_i$
that minimize a positive-definite penalty function
\beq
P = \sum_\ell(\Psi_\ell)^2,
\vspace{-0.2cm}
\eeq
where the $\Psi_\ell$'s are measures of deviation from an ideal fit to
the data.  We use the penalty function described by
\citet{Draine+Allaf-Akbari_2006}, which includes a contribution
proportional to the square of the deviation of the model extinction
curve from the observed extinction curve, and a contribution
proportional to the square of the deviation of the model polarization
curve from the observed polarization curve, as well as terms that
measure the non-smoothness of the size distributions $n_{\sil,i}$ and
$n_{\car,i}$, and of the alignment function $f_i$.  The degree to
which the amount of material in the model grains deviates from the
amount of material ``depleted'' from the gas, as inferred from
observations of ultraviolet absorption lines \citep{Jenkins_2004},
also contributes to the penalty $P$.

\begin{table}[h]
\begin{center}
\caption{\label{tab:models} Models}
\begin{tabular}{ccccc}
\hline
Model        & $b/a$      & $b/a$    & $V_\sil$            &
$V_\car$$^{\rm a}$\\ 
number       & (silicate) & (carbon) & ($10^{-27}\cm^3/$H) &
($10^{-27}\cm^3/$H)\\
\hline
0            & 1.0        & 1.0      & 3.00                & 1.91 \\
1            & 1.4        & 1.0      & 3.81                & 1.93 \\
2            & 1.4        & 1.4      & 3.34                & 1.89 \\
3            & 1.6        & 1.0      & 3.18                & 1.92 \\
4            & 1.6        & 1.6      & 3.20                & 1.89 \\
\hline
``Observed'' &            &          & 2.50                & 1.00 \\
\hline
\end{tabular}

$^{\rm a}$Including $\mathrm{C}/\mathrm{H} = 55$~ppm in PAH material,
with assumed carbon density $2.24~\gm/\cm^{3}$.
\end{center}
\end{table}
\end{widetext}


\section{Size Distributions and Alignment Functions}
\label{sec:size_dist_align_func}

\begin{figure*}
\begin{center}
\epsfig{file=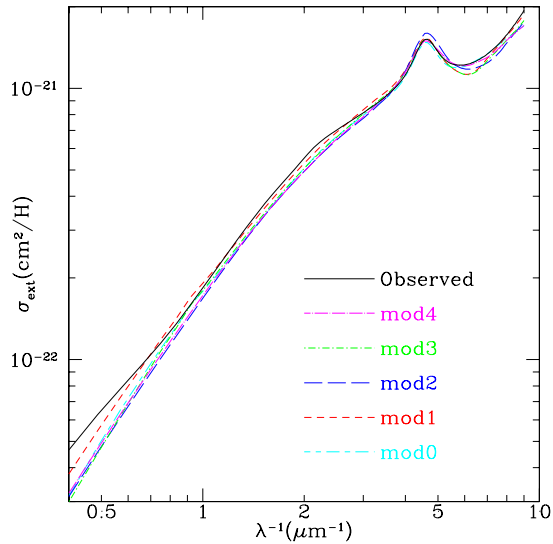,
        width=7.2cm,
	angle=0}
\caption{\label{fig:ext vs lambda} Extinction per H nucleon as a
  function of inverse wavelength for models 0, 1, 2, 3, and 4 (see
  Table~\ref{tab:models}).  The solid line corresponds to the
  ``observed" extinction (as defined in \S~\ref{sec:picket_fence}),
  which appears to be well fit by all five of our models at all
  wavelengths of interest.} 
\end{center}
\end{figure*}

\begin{figure*}
\begin{center}
\epsfig{file=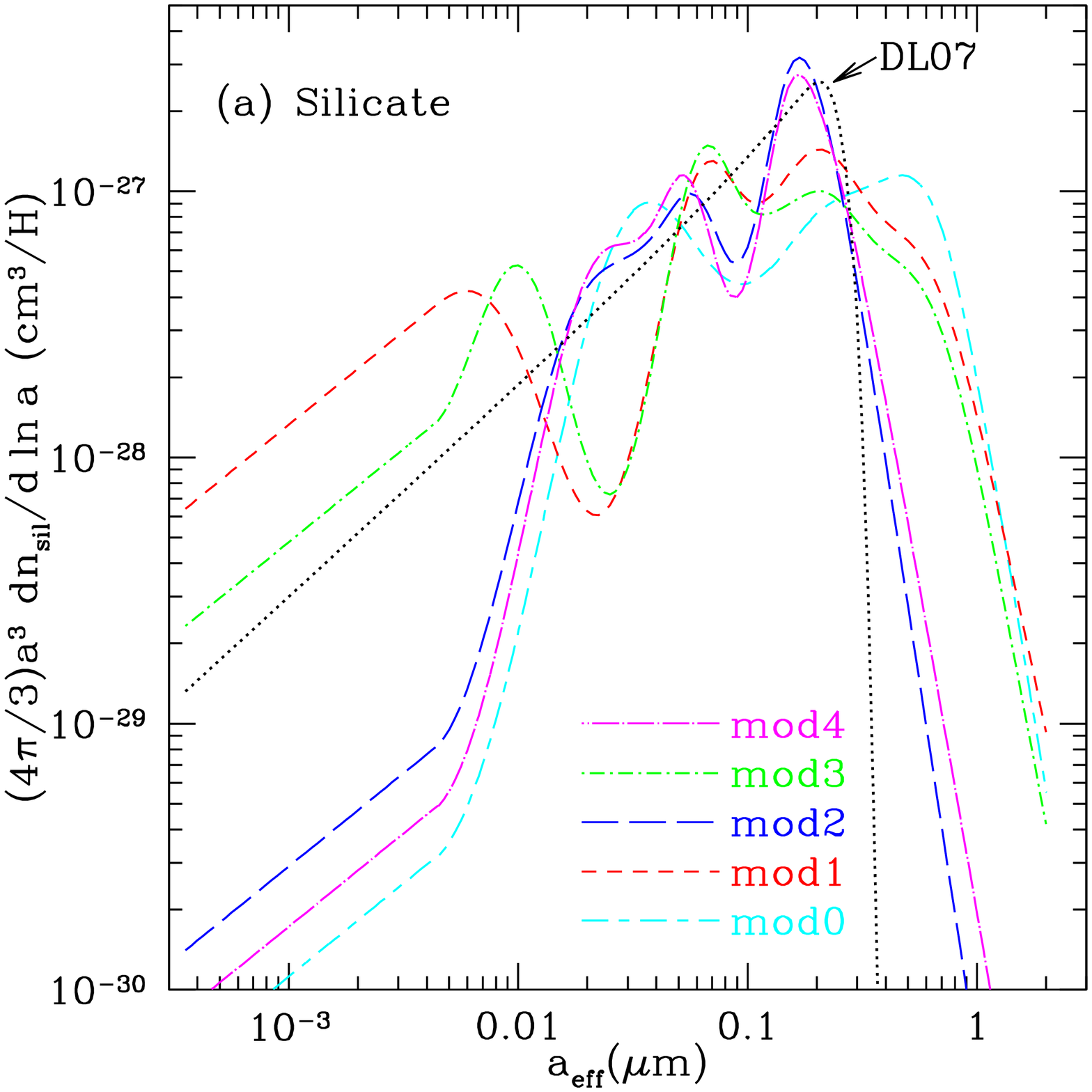,
        width=7.2cm,
	angle=0}
\epsfig{file=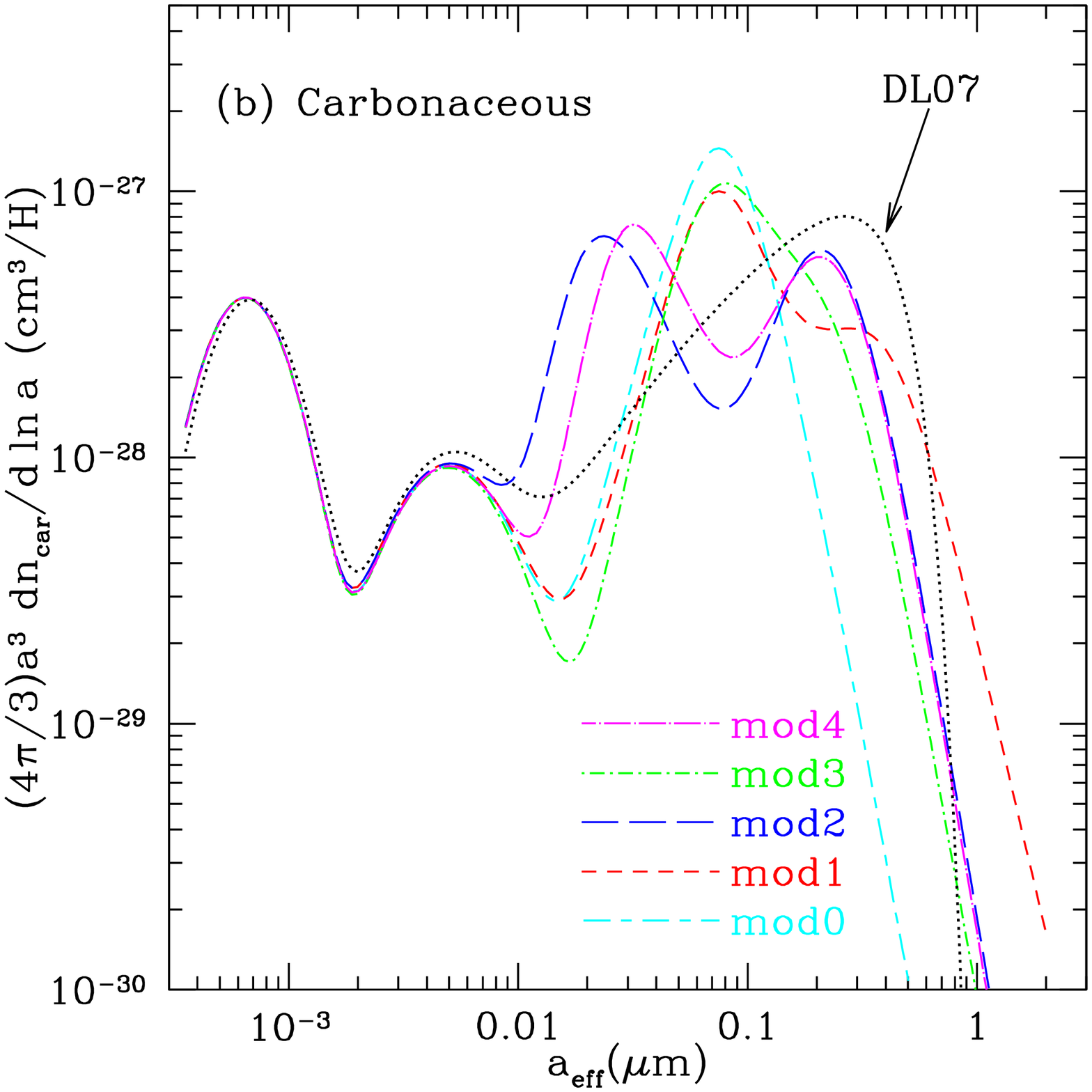,
        width=7.2cm,
        angle=0}
\caption{\label{fig:dmdlna vs a} Mass distribution for ({\it left})
	 silicate and ({\it right}) graphite spheres and spheroids.
	 Results are shown for models 0 through 4.  The curves labeled
	 ``DL07" show the size distributions of
	 \protect\citet{Weingartner+Draine_2001}, as modified by
	 \protect\citet{Draine+Li_2007}.}
\end{center}
\end{figure*}

Polarization in the $9.7~\micron$ silicate feature is routinely
observed \citep{Smith+Wright+Aitken_etal_2000}, and therefore it is
clear that the silicate grains must be nonspherical.  As a result, all
of our models (1 -- 4) include spheroidal silicate grains, with axial
ratios $b/a = 1.4$ or $b/a = 1.6$.  Because there is no observational
evidence of polarization produced by carbonaceous grains, we will
consider two possibilities, namely, that the carbonaceous grains are
spherical, or just as nonspherical (and aligned) as the silicates.

We study four models involving nonspherical grains.  Model 1 has
spheroidal silicate grains with $b/a = 1.4$, and spherical carbon
grains.  Model 2 assumes both silicate and carbon grains to be
spheroids with $b/a = 1.4$.  Model 3 considers silicate spheroids with
a more extreme axial ratio, $b/a = 1.6$, but with spherical carbon
grains.  Finally, our fourth model assumes both silicate and carbon
grains to be spheroids with $b/a = 1.6$.

We obtained the best-fit grain abundances $n_{i,\car}$ and
$n_{i,\sil}$, and degree of alignment $f_i$, for these four distinct
models using the Levenberg-Marquardt algorithm to minimize the penalty
function $P$ discussed in \S\ref{sec:picket_fence}.

Figure \ref{fig:ext vs lambda} shows the ``observed extinction'' that
we seek to reproduce, together with the extinction obtained for each
model.  For comparison purposes, we also carried out a fit to the
extinction using spherical silicate and carbon grains (model 0), with
no consideration of polarization.  Note that all models are
more-or-less equally good at reproducing the observed extinction.  All
of the models tend to underpredict the extinction for
$\lambda \gtrsim 1.2~\micron$, suggesting that perhaps our adopted
composition is insufficiently absorptive at
$\lambda \approx 2~\micron$.  It should be kept in mind, however, that
the ``observed'' extinction is somewhat uncertain at the longer
wavelengths, owing to the increasing difficulty in determining the
extinction per H nucleon when the H column density cannot be
determined directly using Lyman-$\alpha$ absorption
measurements.\footnote{To accurately measure the weaker extinction in
  the infrared, larger dust columns are needed, but the ultraviolet
  extinction is then so strong that Lyman-$\alpha$ measurements become
  infeasible.}
\citet{Nishiyama+Nagata+Kusakabe_etal_2006} find
$A_\lambda \propto \lambda^{-1.99}$ for
$1.25 \leq \lambda \leq 2.2~\um$, steeper than the ``observed"
extinction used here, and closer to the behavior in our models.  The
volumes of silicate and carbonaceous materials required by each model
is given in Table~\ref{tab:models}.

\begin{center}
\it Elemental Abundances
\end{center}

\begin{figure*}
\begin{center}
\epsfig{file=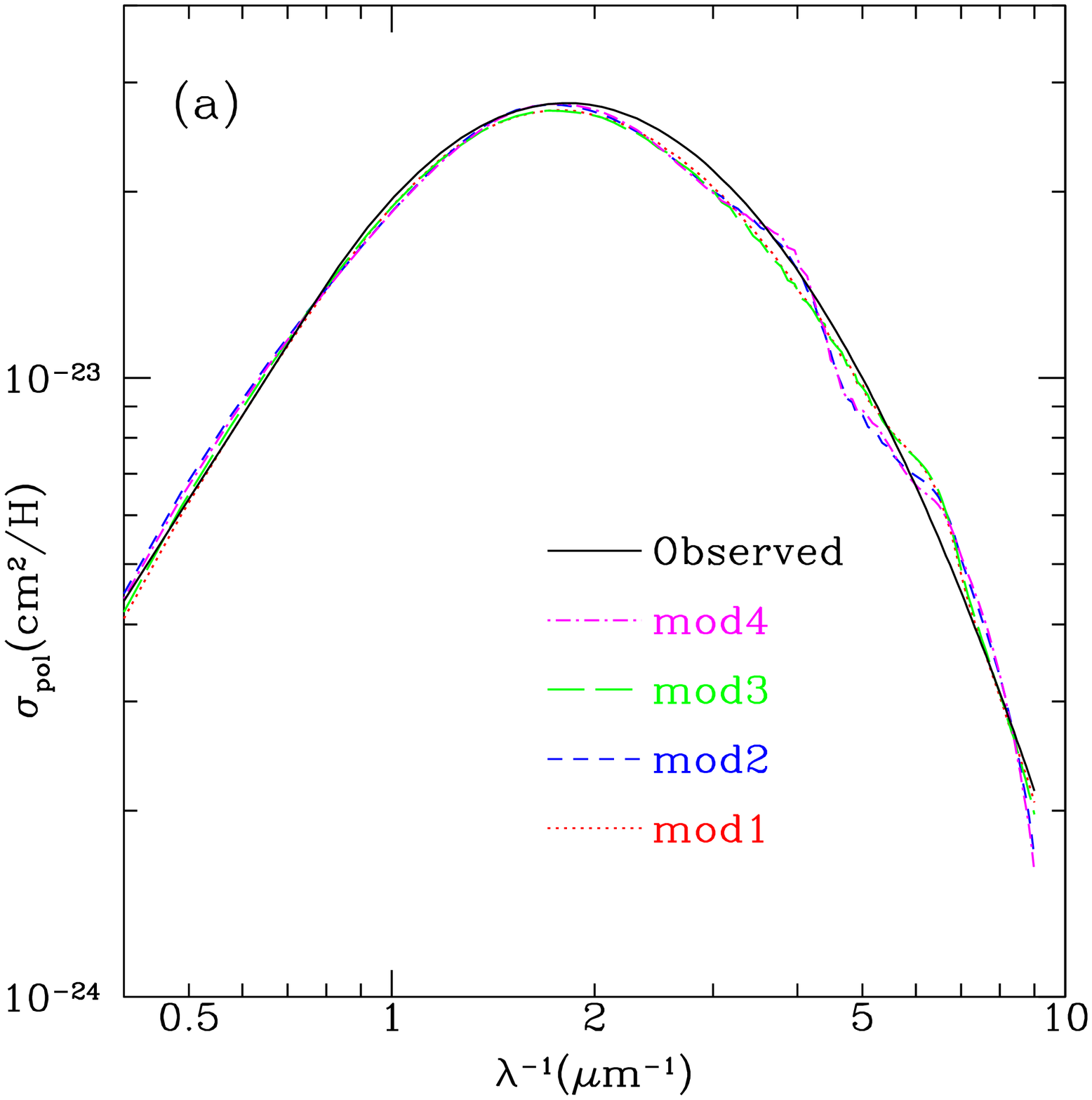,
        width=7.2cm,
	angle=0}
\epsfig{file=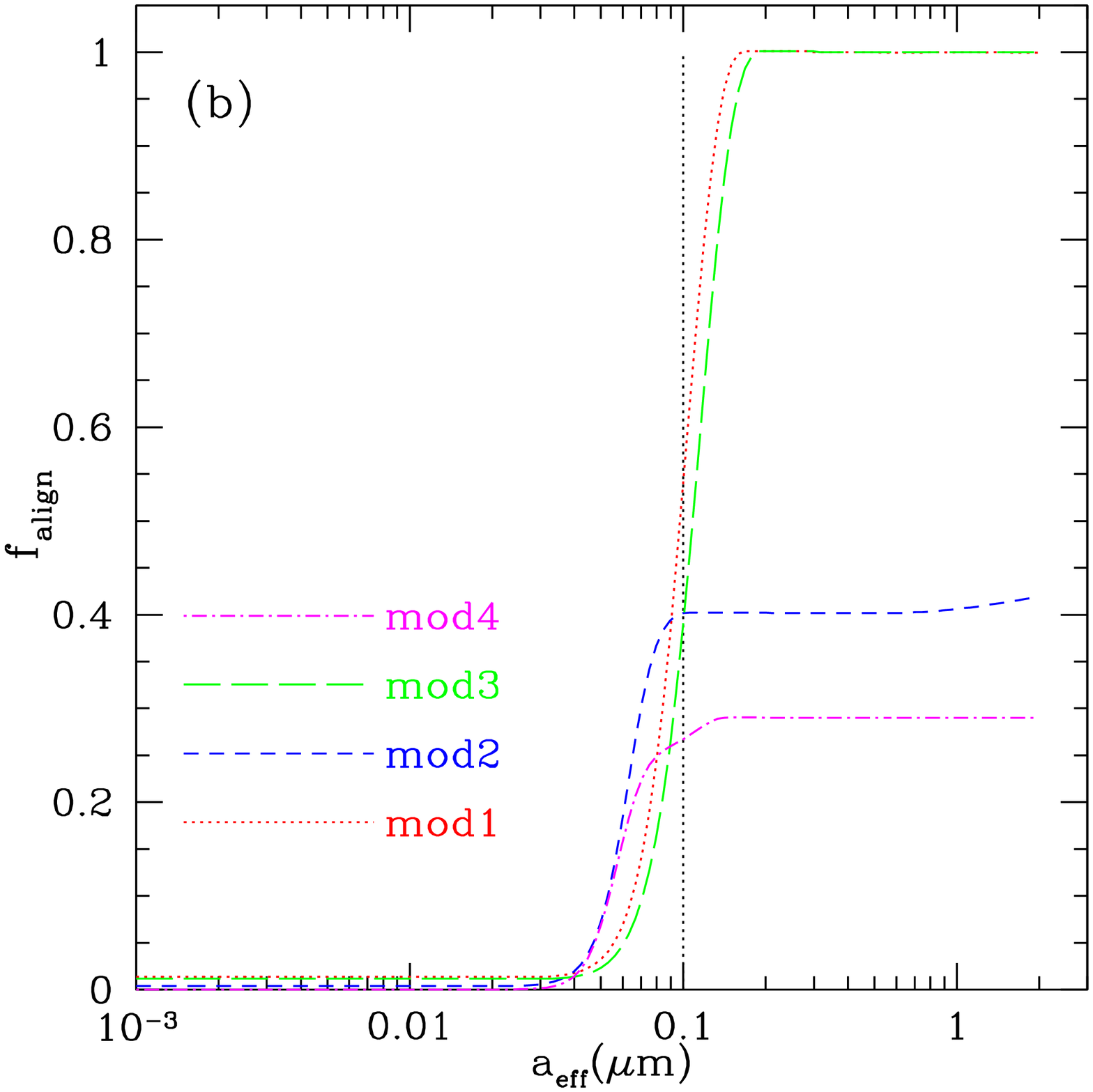,
        width=7.2cm,
	angle=0}
\caption{\label{fig:pol falign} ({\it Left}) Polarization cross
	 section per H nucleon vs. inverse wavelength from
	 observations (solid curve) and for models 1, 2, 3, and 4 (see
	 Table~\ref{tab:models}).  ({\it Right}) Fractional alignment
	 $f$ vs. $a_\eff$ for these four grain models.  The vertical
	 dotted line shows $a_\eff = 0.1~\um$.}
\end{center}
\end{figure*}

The most recent estimates of solar abundances indicate that
${\rm Mg}/{\rm H} = 38$~ppm~\citep{Grevesse+Sauval_1998}, and
${\rm Si}/{\rm H} = 32$~ppm and
${\rm Fe}/{\rm H} = 29$~ppm~\citep{Asplund_2000}.  If all of the Si is
consumed, we could form silicate material with the olivine-like
composition Mg$_{1.1}$Fe$_{0.9}$SiO$_4$.  The density of Si atoms in
fayalite (Fe$_2$SiO$_4$) is about $1.3\times 10^{22}~\cm^{-3}$.  With
this density of Si, ${\rm Mg}_{1.1}{\rm Fe}_{0.9}{\rm SiO}_4$ would
have density $\rho \approx 3.7~\gm/\cm^{3}$, and the solar abundance
of Si could produce $V = 2.5\times 10^{-27}\cm^3/{\rm H}$ of silicate
material.  Thus, we see that the silicate volumes in
Table~\ref{tab:models} exceed what appears to be allowed by current
estimated solar abundances by 20\% to 50\%.

For a carbon density of 2.2~$\gm/\cm^{3}$, the carbon volume in model
0 corresponds to ${\rm C}/{\rm H} = 210$~ppm.  On the well-studied
sightline toward $\zeta$~Oph, the gas-phase carbon abundance is about
$135$~ppm \citep{Cardelli+Mathis+Ebbets_etal_1993}.  Adding this to
our estimate for the carbon in dust grains would give an interstellar
${\rm C}/{\rm H} = 345$~ppm, 40\% greater than the most recent
estimate of the solar abundance
$({\rm C}/{\rm H})_\odot = 245 \pm 30$~ppm
\citep{Asplund+Grevesse+Sauval_etal_2005}.

The amount of grain material called for in these fits is cause for
concern -- our models tend to require about $40\%$ more silicates and
carbon than appears to be available if interstellar abundances are the
same as solar abundances.  However, interstellar abundances of C, Mg,
Si, and Fe may exceed the abundances of these elements in the Sun.
Furthermore, solar abundances may have been underestimated.  For
example, it has recently been argued that the oxygen abundance in the
solar photosphere has been underestimated by a factor of about
2~\citep{Landi+Feldman+Doschek_2007, Centeno+Socas-Navarro_2008}.

~

Figure \ref{fig:dmdlna vs a} shows the size distributions for the
silicate grains and the carbon grains for our five dust models.  Two
classes of models are clearly visible.  The models where only the
silicate grains are aligned (models 1 and~3) have broad distributions
for the silicates, while the carbonaceous grains have a pronounced
peak near $0.08~\um$.  On the other hand, in models where both
silicate and carbonaceous grains are aligned (models 2 and~4), the
silicate size distributions have a dominant peak near $0.15~\um$,
while the carbonaceous grains have pronounced peaks near $0.03~\um$
and $0.25~\um$.  Despite these differences, all of the models 1
through 4 produce polarization that is in reasonable agreement with
the observations, as can be seen in Figure~\ref{fig:pol falign}, but
this is achieved with quite different alignment functions $f(a)$, also
shown in Figure~\ref{fig:pol falign}.  The models with spherical
carbon grains (models 1 and 3) have alignment functions that rise
rapidly from near-zero to near-unity at $a \approx 0.1~\micron$.  When
nonspherical carbon grains are included, however, the large
polarization cross sections for the carbon grains mean that the
required polarization can be achieved without complete alignment of
the $a \gtrsim 0.15~\micron$ grains.  For model 2 ($b/a = 1.4$), the
alignment for $a_\eff \gtrsim 0.1~\micron$ is $f \approx 0.4$; when
$b/a$ is increased to 1.6, the alignment of the larger grains drops to
about $0.3$.  Note that these alignment functions reproduce the
highest observed values of $P_{\rm max}/A_V$.  It is possible that, in
some regions, the dust grains have lower degrees of alignment. 


\begin{widetext}
\vspace{0.001cm}

\section{Far-Infrared Cross Sections and Emissivities}
\label{sec:formalism}

At wavelengths $\lambda \gg a$ (the ``Rayleigh limit''), the
absorption cross section of a grain depends only on the orientation of
the oscillating electric field relative to the grain shape, and not on
the direction of the oscillating magnetic field (i.e., 
$C_{\rm abs,3} = \Cabsb$).  For a spheroid, the absorption cross
section is then 
\beq
\Cabs = \Cabsa\,\cos^2\zeta + \Cabsb\,\sin^2\zeta,
\eeq
where $\zeta$ is the angle between the electric field $\bE$ and the
symmetry axis $\bahat$.  Randomly-orien\-ted spheroids have
$\langle\cos^2\zeta\rangle\nobreak=\nobreak1/3$, which leads to the
``average" absorption cross section 
\beq
\label{eq:abs_cross_sec}
\langle \Cabs \rangle = \frac{1}{3}(\Cabsa + 2\,\Cabsb).
\eeq

Let $\theta_0$ be the angle between the direction of the magnetic
field, $\bxhat$, and the direction of propagation.  We denote
quantities for radiation with $\bE$ parallel (respectively,
perpendicular) to  the $(\bk,\bxhat)$ plane with the label $\parallel$
(respectively,~$\perp$).  If a fraction $f$ of the grains is perfectly
aligned, and a fraction $(1 - f)$ is randomly-oriented, 
\beqa
\left<\cos^2\zeta_{\hspace{0.04cm}\parallel\hspace{0.04cm}}\right> & 
      = & \frac{1 - f}{3} + f\sin^2\theta_0,\\
\left<\cos^2\zeta_\perp\right> & = & \frac{1 - f}{3}.
\eeqa
The polarization-averaged absorption cross section is then given by
\beq
\Cabs \equiv
\frac{1}{2}\,\big(\Cabs^\perp + \Cabs^\parallel\big) =
\frac{1}{2}\left\{
   \Cabsa \left[
      \frac{2}{3} - f\left(\cos^2\theta_0 - \frac{1}{3}\right)\right] +
   \Cabsb\left[
      \frac{4}{3} + f\left(\cos^2\theta_0 - \frac{1}{3}\right)\right]
\right\}\!,
\eeq
whereas the cross section for polarized emission is defined as
\beq
\Cpol \equiv
\frac{1}{2}\,\big(\Cabs^\perp - \Cabs^\parallel\big) =
\frac{1}{2}\,f\left(\Cabsb - \Cabsa\right)\sin^2\theta_0.
\eeq

Therefore, for a mixture of partially-aligned grains, the total and
polarized emissivities are  
\beqa \nonumber
\frac{1}{N_{\rm H}}\,I_\lambda(\theta_0) & = &
\frac{1}{3}\,\sum_{m}
\int \dd a_\eff\, \frac{\dd n_m}{\dd a_\eff}\,
(2\,\Cabsb + \Cabsa) \left<B_\lambda\right>_{m,a_\eff} \\
                                         & + &
\frac{1}{2}\left(\cos^2\theta_0 - \frac{1}{3}\right) \sum_{m}
\int \dd a_\eff \,\frac{\dd n_m}{\dd a_\eff} f(a_\eff)
(\Cabsb - \Cabsa) \left<B_\lambda\right>_{m,a_\eff}\!, \\
\frac{1}{N_{\rm H}}\,I_{\lambda,{\rm pol}}(\theta_0) & = &
\frac{1}{2}\,\sin^2\theta_0 \sum_{m}
\int \dd a_\eff\, \frac{\dd n_m}{\dd a_\eff}\, f(a_\eff)
(\Cabsb - \Cabsa) \left<B_\lambda\right>_{m,a_\eff}\!,
\eeqa
where
\beq
\left<B_\lambda\right>_{m,a_\eff} \equiv
\int \dd T\, \frac{\dd P(a_\eff,m,U)}{\dd T}\,B_\lambda(T),
\eeq
$\dd P(a_\eff,m,U)/\dd T$ is the temperature distribution for grains
of type $m$ and size $a_\eff$ in a radiation field of intensity~$U$,
and $B_\lambda(T)$ the Planck function.  We use the temperature
distributions computed by~\citet{Draine+Li_2007} for spherical
grains.  For the spheroidal components of our models, we correct these
distributions to account for the non-sphericity of the grains by
requiring that the power they absorb equals the power they re-radiate.
The correction is very small. 
\vspace{0.4cm}  
\end{widetext}


\section{Results and Discussion}
\label{sec:ir_em}

\begin{figure*}
\begin{center}
\epsfig{file=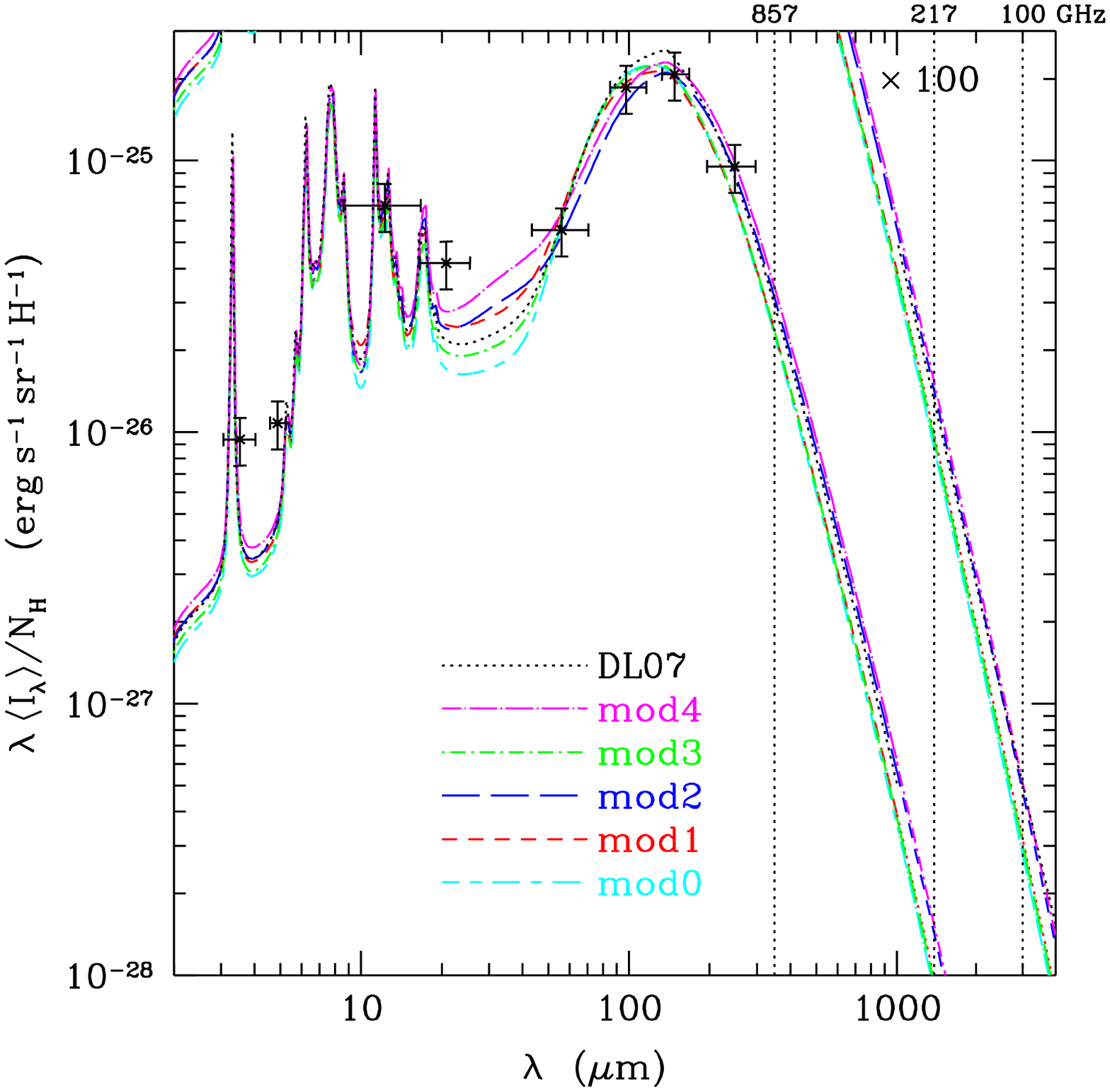,
        width=7.93cm,
	angle=0}
\caption{\label{fig:em_spectra} Total emission spectra for the
	 different grain models listed in Table~\ref{tab:models}, as
	 well as for the model described
	 in~\protect\citet{Draine+Li_2007}, each heated by the
	 interstellar radiation field estimated 
	 by~\protect\citet{Mathis+Mezger+Panagia_1983}.  The $x$-axis
	 is now labeled by wavelengths, and the symbols correspond to
	 the DIRBE results for HI correlated emission at high Galactic
	 latitudes~\protect\citep{Dwek+Arendt+Fixsen_etal_1997,
	 Arendt+Odegard+Weiland_etal_1998}.  Frequencies covering the
	 full range of \emph{Planck} HFI frequencies (100~GHz, 
	 217~GHz, and 857~GHz) are also indicated.}
\end{center}
\end{figure*}

Figure \ref{fig:em_spectra} shows the emission spectra calculated for
each of the five grain models listed in Table~\ref{tab:models} when
illuminated by the standard interstellar radiation field for the solar
neighborhood.  Also shown is the emission spectrum calculated for the
grain model of~\citet{Draine+Li_2007}.  The emission
$\left<I_\lambda\right>$ is calculated using the cross section
$\left<\Cabs\right>$ from equation~(\ref{eq:abs_cross_sec}),
appropriate for randomly-oriented dust.  These spectra of the thermal
emission do not include rotational emission from spinning dust, which
appears to dominate for $\lambda \gtrsim 4000~\um$
($\nu \lesssim 75$~GHz) \citep{Draine+Lazarian_1998,Finkbeiner_2004}.

\begin{figure*}
\begin{center}
\epsfig{file=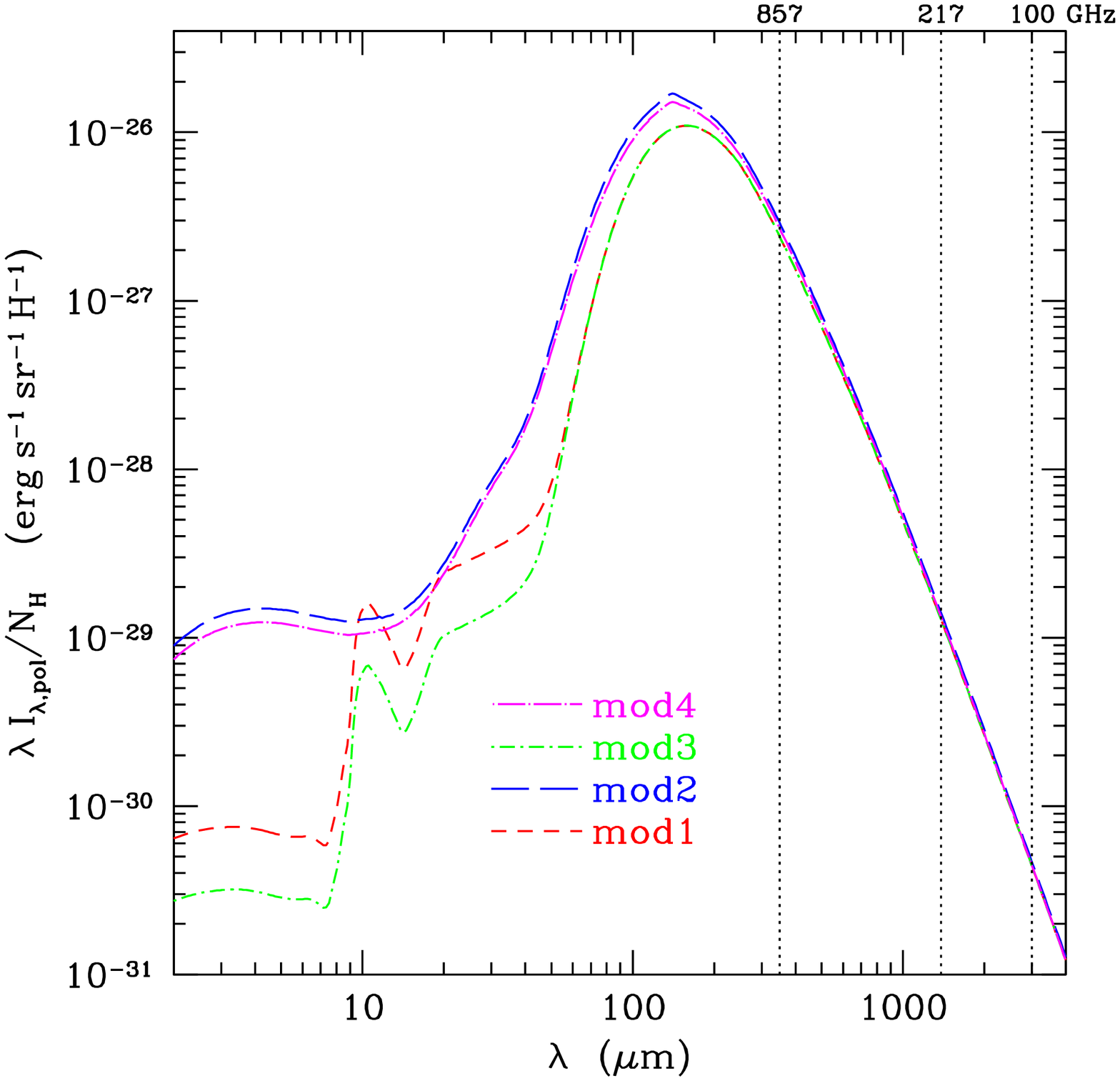,
        width=7.93cm,
        angle=0}
\hspace*{-0.5cm}
\epsfig{file=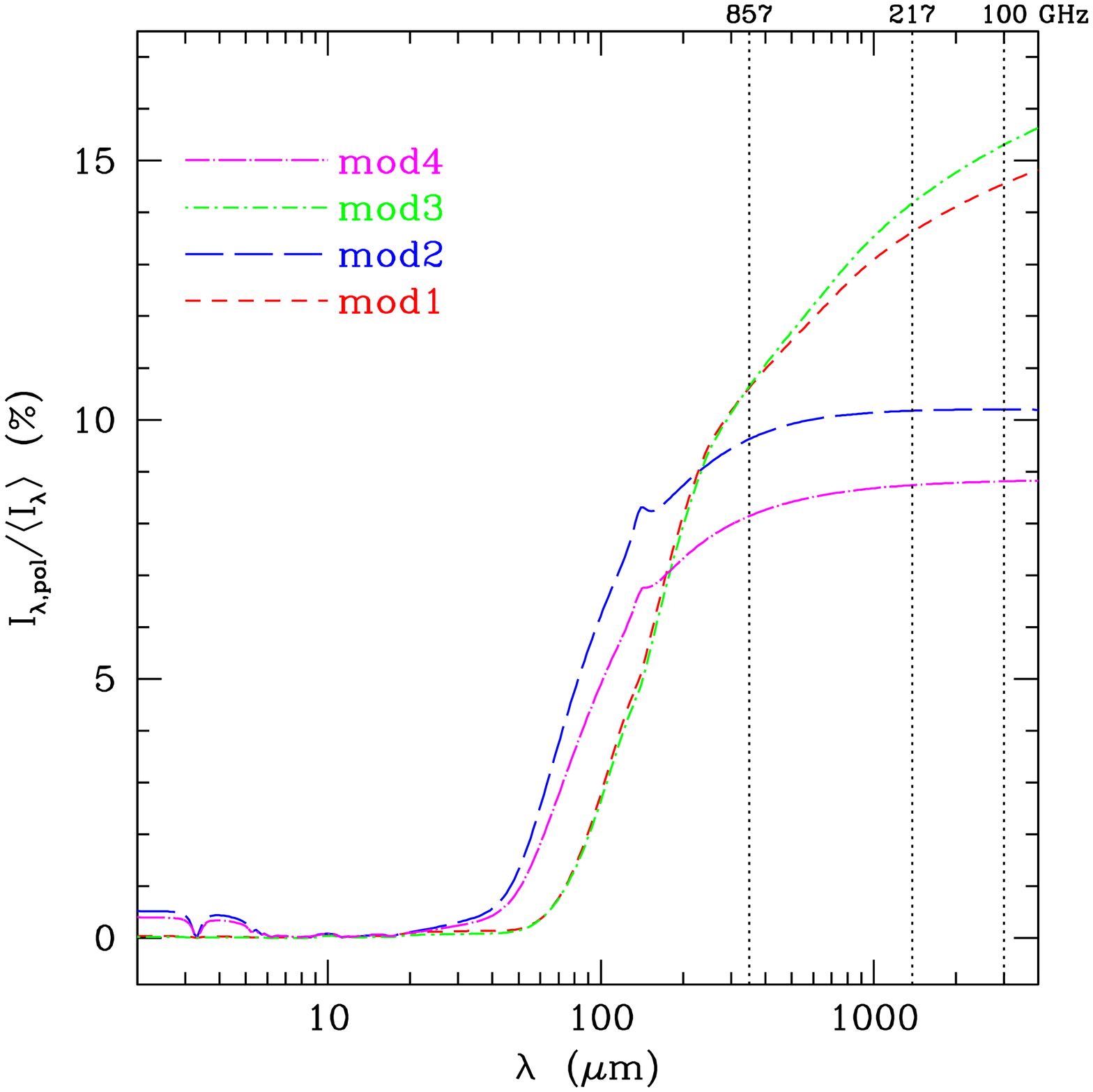,
        width=7.93cm,
        angle=0}
\caption{\label{fig:pol_spectra} ({\it Left}) Spectra of polarized
         thermal emission predicted by models 1 through 4  in
         directions perpendicular to the direction of the local
         magnetic field, and ({\it Right}) corresponding degrees of
         polarization for each model. Two classes of models are
         clearly visible.  Note that these models do not include
         rotational emission from spinning dust, which will be
         important for $\lambda \gtrsim 4000~\um$
	 ($\nu \lesssim 75$~GHz).}
\end{center}
\end{figure*}

The emission spectra for all six models are relatively similar.  In each case,
$\lambda\,\langle I_\lambda \rangle$ peaks near $130~\micron$, with an
amplitude within $20\%$ of the~\citet{Draine+Li_2007} (thereafter,
DL07) model.  Moreover, all models agree extremely well for
wavelengths below about $20~\um$, where the signal is entirely
dominated by PAH features. 

However, there are some noticeable differences, especially in the
ratio of the 130~$\um$ to the 24~$\um$ emissions.  Although these
ratios are within $15\%$ of each other for model 0 and the DL07 model,
models involving spheroidal grains lead to smaller values of this
ratio, typically by a factor of 1.5.  This difference arises mostly
from variations in the size distributions between models, with
differences in grain shape also having a small effect on the average
emission spectrum.  All other differences are fairly modest (usually
at the 10\% to 20\% level), and generally smaller than the actual
uncertainties in the appropriate dielectric functions.

The fact that all of these spectra are in good agreement overall, but
with significant well localized differences, opens a very interesting
window through which one can hope to strongly constrain dust models
with observations of the total dust emission at just a couple of
wavelengths.  We have not attempted this exercise here.

Figure~\ref{fig:pol_spectra} shows the polarized emission spectra
calculated for models 1 through 4, as well as the corresponding
degrees of polarization (in \%), for lines of sight perpendicular to
the direction of the local magnetic field, and the dust exhibiting the
highest degrees of alignment found in the diffuse interstellar
medium. 

Contrary to the total emission spectra shown on
Figure~\ref{fig:em_spectra}, the polarized emission spectra clearly
reflects the fact that we consider two classes of models.  The spectra
of polarized emission for models 2 and 4, with spheroidal silicate and
carbonaceous grains, are indeed remarkably similar, but quite
different from those for models 1 and 3, with silicate spheroids and
carbonaceous spheres.  This translates into interestingly different
predictions for the expected degree of polarization for emission by
interstellar dust, as can be seen in the right panel of
Figure~\ref{fig:pol_spectra}. 

Although all models predict virtually no polarization below $40~\um$
(as small grains and PAHs are not aligned), the wavelength dependence
of the degree of polarization for $\lambda \gtrsim 40~\um$ is quite
model-dependent.  Models with both silicate and carbonaceous
spheroidal grains lead to the degree of polarization sharply
increasing between about $40~\um$ and $600~\um$, after which it
reaches a plateau.  However, the degree of polarization for models in
which only the silicate grains are spheroids increases steadily with
increasing wavelength, without converging to an upper limit.  As a
result, models 1 and 3 predict significantly higher degrees of
polarization than models 2 and 4 for wavelengths longer than about
$200~\um$.  For example, when $\lambda = 3000~\um$, the maximum degree
of polarization predicted by models 1 and 3 is of order $10\%$, while
models 2 and 4 lead to a maximum of about $15\%$.  These differences
provide another way of discriminating between dust models, this time
from observations of the polarized emission from interstellar dust.

The strong wavelength-dependence of the polarization in models 1 and 3
can be easily understood.  In these two models, the carbonaceous
grains are not aligned.  For the adopted grain optical properties, the
silicate grains contribute an increasing fraction of the emission as
the wavelength increases, in part because the silicate grains are
slightly cooler than the carbonaceous grains (this explains the
behavior at $\lambda \lesssim 200~\um$), and in part because the ratio
of the silicate opacity to the graphite opacity increases with
increasing wavelength for $\lambda \gtrsim 100~\um$. 


\section{Summary and Further Discussion}
\label{sec:summary}

The main results of this paper are as follows:
\begin{enumerate}
\item We present models of interstellar dust based on mixtures of
  spheroidal amorphous silicate grains, spheroidal graphite grains,
  and PAH particles, with a degree of alignment $f$ that is a function
  of the effective radius $a_\eff$ of the grains.  The models differ
  in assumptions concerning the shapes (axial ratios $b/a = 1.4$ and
  1.6 are considered for the spheroidal grains) of the silicate and
  graphite grains. 
\item Size distributions $\dd n/\dd a_\eff$ and alignment function
  $f(a_\eff)$ are found that reproduce the observed
  wavelength-dependent extinction and polarization of starlight.  As
  found previously by \citet{Kim+Martin_1995}, the degree of alignment
  is small for $a_\eff\nobreak\lesssim\nobreak0.05~\micron$, and large
  for $a_\eff\nobreak\gtrsim\nobreak0.1~\micron$.
\item For models where only the silicate grains are aligned, the
  degree of alignment approaches $100\%$ for $a \gtrsim 0.1~\micron$,
  whereas when both silicate and graphite grains are partially-aligned
  spheroids with $b/a = 1.4$, the observed starlight polarization can
  be reproduced with $f \approx 0.4$ for $a \gtrsim 0.1~\micron$.  If
  the axial ratio is increased to $b/a = 1.6$, the required degree of
  alignment drops to $f \approx 0.3$.
\item For each model, we calculate the partially-polarized infrared
  emission when the dust is illuminated by the local starlight
  background.  The total infrared emission spectra are generally
  similar, and in approximate agreement with observations of the
  Galactic cirrus emission at $|b| \gtrsim 20^\circ$.  In detail, the
  spectra are model-dependent, particularly between about $20~\um$ and
  $160~\um$.  No attempt has been made to adjust our models to
  reproduce observed FIR spectral energy distributions.
\item Because the small grains and PAHs are not aligned, the polarized
  emission is very weak for $\lambda \lesssim 40~\micron$.  However,
  for $\lambda \gtrsim 60~\micron$, the linear polarization of the
  thermal emission is appreciable when viewed from directions
  perpendicular to the local magnetic field.  At $100~\micron$, the
  different models considered here result in linear polarizations
  between about $3\%$ and $6\%$.  For wavelengths longer than
  $500~\um$ ($\nu \leq 600$~GHz), of relevance to CMB studies, the
  degrees of polarization predicted by our models range from about
  $9\%$ to around $15\%$ for the thermal emission.  However, for
  $\lambda \gtrsim 4000~\um$ ($\nu\nobreak\lesssim\nobreak75$~GHz),
  one should take into account the additional emission from spinning
  dust, which is not included here.  This rotational emission is
  expected to be nearly unpolarized~\citep{Lazarian+Draine_2000},
  which will lower the overall polarization. 
\item For models that employ more than one type of grain, which is the
  case of all our models, their relative contribution to the
  far-infrared and submillimeter emission will in general vary with
  wavelength.  If the two types of grain have different values of
  $C_{\rm pol}/C_{\rm abs}$ because of different alignment fractions
  or grain properties, the degree of polarization will vary with
  wavelength.  This effect is very pronounced in our models 1 and 3
  where only the silicate grains are aligned with the magnetic field.
\end{enumerate}

These last two results are particularly interesting for studies of the
polarized components of the CMB.  Polarized dust emission is indeed
one of the brightest Galactic foregrounds at microwave frequencies,
and being able to remove it from Galactic maps is one of the biggest
challenges that will limit our ability to detect B-modes.

The maximum degrees of polarization we calculated in \S\ref{sec:ir_em}
are consistent with the level inferred from the {\it WMAP}
observations~\citep{Page+Hinshaw+Komatsu_etal_2007,
                    Gold+Bennett+Hill_etal_2008},
although $5\%$ is on the lower side of our predictions.  Upcoming
experiments, such as {\it Planck}, might therefore want to allow for
the possibility of dust polarization being slightly higher in their
sky models, and consider how this would affect their ability to
constrain E- and B-modes.

More important is the possible wavelength-dependence of the degree of
polarized dust emission.  Models involving both silicate and
carbonaceous spheroidal grains lead to a roughly constant level of
polarization at all {\it Planck} frequencies.  However, there is
currently no evidence that carbonaceous dust grains contribute to the
polarized signal at any wavelength, and models where only silicate
grains are aligned spheroids lead to degrees of polarization that are
strongly frequency-dependent.  Although a possible challenge for the
CMB community, this will allow the use of Planck observations to test
whether the submm emission arises from a single kind of grain or from
multiple grain types having different values of $\Cpol/\Cabs$. 


\acknowledgements
This work was supported in part by NSF grant AST-0406883.  AAF was
additionally supported by Princeton University and NSF grant
AST-0707932.  The authors thank R.~H.~Lupton for making the SM package
available to them.  This research has made use of NASA's Astrophysics
Data System Bibliographic~Services. 


\bibliographystyle{hapj} 
\bibliography{refs}


\end{document}